\pdfoutput=1
\documentclass[journal]{IEEEtran}
\IEEEoverridecommandlockouts
\usepackage{color}
\usepackage{subfigure}
\usepackage[ruled,vlined,algo2e]{algorithm2e}
\usepackage{algorithm}
\usepackage{algorithmic}
\usepackage{graphicx}
\usepackage{amssymb}
\usepackage{amsmath}
\usepackage[numbers,sort&compress]{natbib}
\usepackage{booktabs}
\usepackage{multirow}
\usepackage{hyperref}
\usepackage[utf8]{inputenc}
\usepackage{xcolor}
\usepackage{booktabs}
\usepackage{array}
\usepackage{pifont}
\usepackage{threeparttable}

\title{Foundation Models for Wireless Communications: \\
From PHY Intelligence to Network Autonomy}

\author{
    Le~Liang,~\IEEEmembership{Member,~IEEE}, 
    Jiajia~Guo,~\IEEEmembership{Member,~IEEE}, 
    Jun~Zhang,~\IEEEmembership{Fellow,~IEEE}, 
    Chan-Byoung~Chae,~\IEEEmembership{Fellow,~IEEE}, 
    Lu~Lu,~\IEEEmembership{Member,~IEEE}, 
    Shugong~Xu,~\IEEEmembership{Fellow,~IEEE}, 
    Octavia~A.~Dobre,~\IEEEmembership{Fellow,~IEEE}, 
    Shi~Jin,~\IEEEmembership{Fellow,~IEEE}, 
    and~Geoffrey~Ye~Li,~\IEEEmembership{Fellow,~IEEE} 

\thanks{L. Liang and S. Jin are with the School of Information Science and Engineering, Southeast University, Nanjing 211189, China. L. Liang is also with the Purple Mountain Laboratories, Nanjing 211111, China. (e-mail: \{lliang, jinshi\}@seu.edu.cn).}

\thanks{J. Guo and J. Zhang are with the Department of Electronic and Computer Engineering, Hong Kong University of Science and Technology, Hong Kong, China (e-mail: \{eejiajiaguo, eejzhang\}@ust.hk).}

\thanks{C.-B. Chae is with the School of Integrated Technology, Yonsei University, Seoul 03722, South Korea (e-mail: cbchae@yonsei.ac.kr).}

\thanks{L. Lu is with MediaTek, USA (e-mail: Lu.Lu@mediatek.com).}

\thanks{S. Xu is with Xi’an Jiaotong-Liverpool University, Suzhou 215123, China (e-mail: shugong.xu@xjtlu.edu.cn).}

\thanks{O. A. Dobre is with the Faculty of Engineering and Applied Science, Memorial University of Newfoundland, St. John's, NL A1B 3X5, Canada (e-mail: odobre@mun.ca).}

\thanks{G. Y. Li is with the Department of Electrical and Electronic Engineering, Imperial College, London SW7 2AZ, U.K. (e-mail: geoffrey.li@imperial.ac.uk).}
}

\date{}

\begin{document}

\maketitle

\begin{abstract}
6G networks will introduce unprecedented complexity, which calls for a paradigm shift in network optimization and management. Artificial intelligence (AI)-based solutions, especially those enabled by the recently developed foundation models, have been recognized as promising candidates. Foundation models are large-scale AI models with general-purpose feature extraction capabilities, and once trained on massive amounts of data, they can be adapted to solve a wide range of downstream tasks, either in a zero-shot manner or with few-shot fine-tuning. This article provides a comprehensive overview of how foundation models are reshaping physical-layer processing and wireless resource management across three progressive paradigms. First, we examine the adaptation of off-the-shelf pre-trained foundation models to various wireless tasks. Second, we explore wireless-native foundation models, built from scratch on wireless data to bridge cross-domain modality gaps and capture universal wireless-domain physical characteristics. Third, we highlight agentic foundation models, which elevate static data processing into autonomous, reasoning-driven network orchestration. Furthermore, we discuss the impact of applying foundation models to emerging 6G frontiers, including integrated sensing and communications (ISAC), new multiple-input multiple-output (MIMO) architectures, semantic communications, and system-level network autonomy. Finally, we identify critical open challenges and opportunities, charting a promising path toward fully intelligent and adaptive wireless networks.
\end{abstract}

\begin{IEEEkeywords}
Wireless foundation models, 6G, large language models, agentic artificial intelligence, physical layer, resource management, integrated sensing and communications, semantic communications, MIMO systems.
\end{IEEEkeywords}

\section{Introduction}

\IEEEPARstart{O}ver the past decade, the rapid evolution of artificial intelligence (AI) has catalyzed a profound revolution in wireless communications. By shifting away from rigid mathematical models and heuristic algorithms, modern AI-driven solutions harness massive volumes of wireless data to train neural networks (NNs) capable of learning intrinsic environmental patterns. This convergence of AI and wireless technologies has established a fundamentally new paradigm, elevating performance capabilities across physical-layer processing and wireless resource management. At the physical layer, such AI-driven approaches demonstrate exceptional efficacy in tasks ranging from channel state information (CSI) acquisition to transceiver design. Furthermore, intelligent algorithms have profoundly transformed wireless resource orchestration and autonomous network management, establishing AI as an indispensable pillar of 6G networks.

Despite these successes, conventional task-specific NNs face severe bottlenecks as wireless networks evolve toward the 6G era, which will be characterized by unprecedented complexity and more stringent and heterogeneous service requirements. First, such traditional AI-based methods often suffer from poor generalization, experiencing severe performance degradation when deployed in unseen environments with different channel statistics. Second, their limited representational capacity imposes a strict performance ceiling, restricting their ability to capture high-dimensional wireless patterns. Finally, as dedicated models must be trained for each functional block in communication systems, they fundamentally fail to support multi-task operations, resulting in profound inflexibility. To overcome these bottlenecks, foundation models have recently emerged as a promising solution, first revolutionizing natural language processing (NLP) and computer vision areas, then extending their growing impact on wireless communications and other scientific fields. By definition, a foundation model is a massive neural architecture pre-trained on vast datasets using self-supervised learning, designed to be adapted to a wide array of downstream tasks. Unlike narrow NNs built from scratch for one specific objective, foundation models extract universal representations. This endows them with powerful general-purpose features, enabling remarkable zero-shot or few-shot generalization. The transformative power of this paradigm was exemplified by large language models (LLMs)~\cite{devlin2019bert,radford2019language}, as well as large vision models (LVMs)~\cite{dosovitskiy2021image} and multi-modal architectures. Recent advances in LLM-enabled wireless communications have further revealed a transition from task-level adaptation to autonomous reasoning and coordination \cite{liang2026large}.

Driven by this unprecedented cross-domain versatility, the integration of foundation models into wireless communications follows a three-stage progressive trajectory to conquer the complexities of next-generation networks. Initially, researchers focus on adapting existing pre-trained foundation models, typically LLMs, directly to communication tasks. By treating wireless data as specialized sequences, off-the-shelf pre-trained foundation models are fine-tuned to perform functions such as channel prediction \cite{liu2024llm4cp} and beam prediction \cite{Sheng2025beam}. However, a fundamental modality gap exists between the discrete textual space of natural language and the continuous wireless signals, alongside strict latency constraints that standard autoregressive models struggle to meet. These mismatches necessitate a paradigm shift toward developing wireless-native foundation models. Pre-trained from scratch on massive wireless datasets, wireless-native foundation models bypass artificial token alignment to directly capture wireless characteristics \cite{liu2025wifo,sheng2025wireless}, thus often more compact and lightweight. Channel foundation models (CFMs) \cite{jiang2025towardscfm} are one realization of this paradigm: they are pre-trained on large-scale channel information, such as CSI, channel impulse response (CIR), in-phase and quadrature (I/Q) samples, and received waveforms, and then adapted across wireless tasks, scenarios, and system configurations. However, both the adapted pre-trained foundation models and the wireless-native foundation models primarily function as static feature encoders lacking proactive intelligence. To overcome this limitation, agentic foundation models have recently emerged \cite{Shao2024wirelessllm,Fan2025mapc}. By empowering neural architectures with domain knowledge, logical reasoning loops, and external tool use, agentic foundation models elevate the physical-layer processing and wireless resource management into autonomous systems capable of actively perceiving dynamic environments, facilitating multi-agent collaboration, and executing complex network orchestration.

The main contributions of this article are summarized as follows:
\begin{itemize}

    \item We present a unified taxonomy of wireless foundation models across three evolutionary paradigms: adapted pre-trained models, wireless-native models, and agentic models for autonomous decision-making.

    \item We provide a comprehensive tutorial-style overview of how foundation models are being applied to physical-layer processing, including CSI acquisition, CSI feedback, channel prediction, signal detection, beam prediction, and multi-task signal processing.

    \item We extend the discussion from signal-level intelligence to network-level intelligence by reviewing the role of foundation models in wireless resource allocation, interference management, intent-driven orchestration, and autonomous network control.

    \item We highlight emerging application domains in next-generation wireless systems, including integrated sensing and communications, new multiple-input multiple-output (MIMO) architectures, semantic communications, and system-level network autonomy.

    \item We identify key open challenges and future research directions, including wireless-native pre-training, real-time deployment, dataset and benchmark construction, trustworthy agentic reasoning, and cross-layer foundation-model design.
\end{itemize}

To provide a comprehensive understanding of this transformative technology, this article presents an overview of foundation models in wireless communications. Section~\ref{sec:basics} outlines the fundamental principles of foundation models. Section~\ref{sec:PHY} explores their integration into physical-layer processing, detailing the evolutionary progression from adapting pre-trained foundation models, to wireless-native foundation models, and ultimately to agentic foundation models, alongside a review of essential datasets. Section~\ref{sec:RA} extends this progressive framework to wireless resource management, examining intelligent multi-user resource allocation and intent-driven resource orchestration. Section~\ref{sec:Emerging_Technologies} highlights emerging 6G applications, including integrated sensing and communications (ISAC), fluid antenna and new MIMO systems, semantic communications, and system-level network autonomy. Finally, Section~\ref{sec:challenges} concludes by identifying critical open challenges and future research directions.

\begin{table*}[t]
\caption{Representative Foundation Model Families in General AI}
\label{tab:ai-fm-examples}
\centering
\small
\setlength{\tabcolsep}{6pt}
\renewcommand{\arraystretch}{1.1}
\begin{tabular}{
>{\raggedright\arraybackslash}p{0.16\textwidth}
>{\raggedright\arraybackslash}p{0.28\textwidth}
>{\raggedright\arraybackslash}p{0.47\textwidth}
}
\toprule
\textbf{Family} & \textbf{Representative models} & \textbf{Core structure and use} \\
\midrule
LLMs 
& GPT-series, LLaMA-family 
& Mainly decoder-only Transformer; strong text generation, reasoning, and instruction following~\cite{brown2020language}. \\

LVMs 
& ViT, masked autoencoder (MAE)-style models 
& Mainly encoder-only Transformer over patch tokens; transferable visual representation learning~\cite{dosovitskiy2021image}. \\

Multi-modal models 
& VLM/multi-modal LLM families (e.g., CLIP-like, LLaVA-like) 
& Cross-modal fusion/alignment between language and vision backbones for joint understanding and generation~\cite{li2025survey}. \\
\bottomrule
\end{tabular}
\end{table*}

\section{Basics of Foundation Models}
\label{sec:basics}
This section provides a brief introduction to the fundamental principles of foundation models to set the stage for subsequent discussions on wireless foundation models. 

\begin{figure}
\centering
\includegraphics[width=0.98\linewidth]{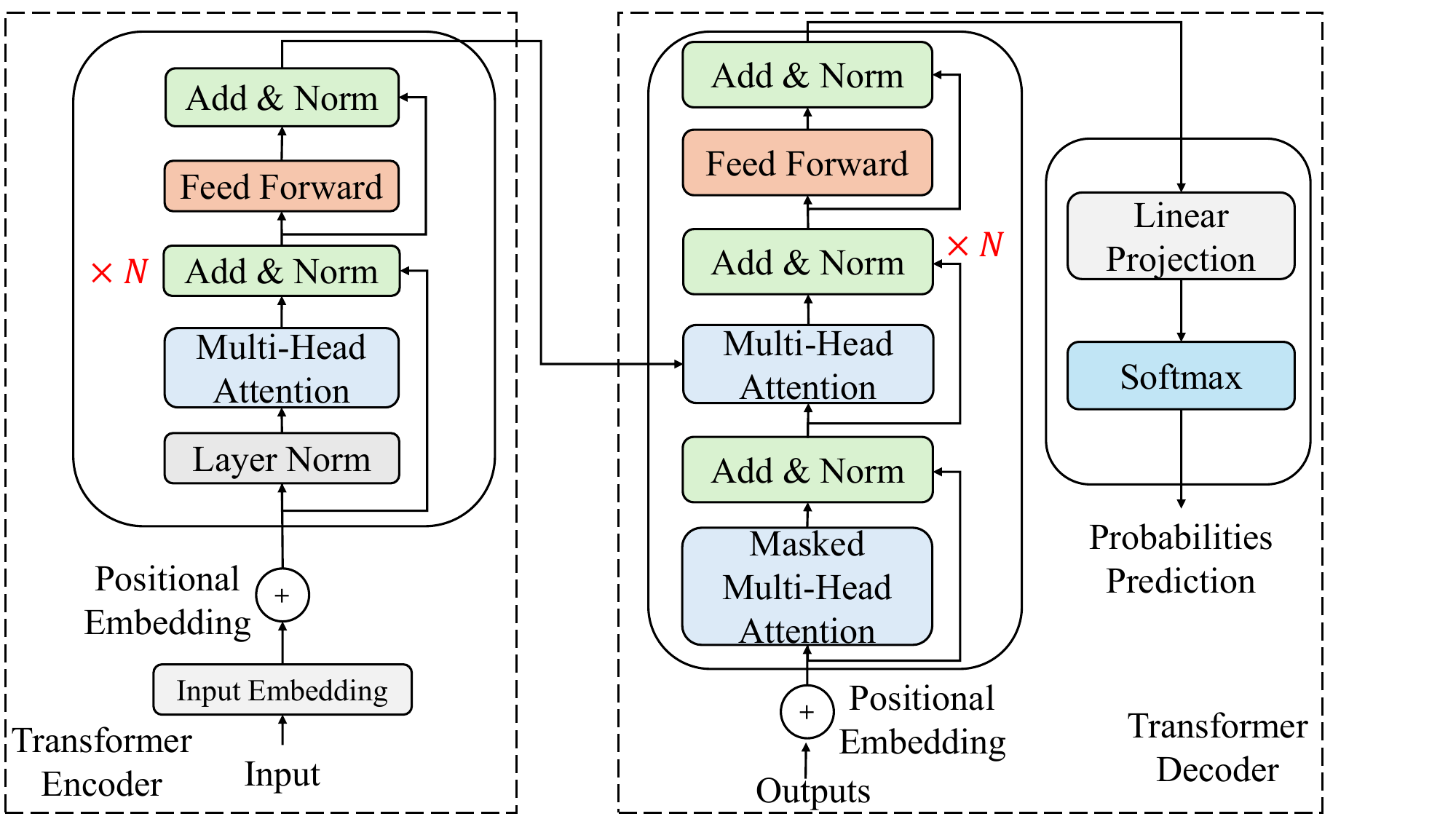}
\caption{Architectural illustration of encoder-decoder and decoder-only Transformers. The encoder-decoder configuration is a versatile framework for sequence transduction, whereas the decoder-only architecture serves as the prevailing backbone for modern generative LLMs.}
\label{fig:transformer-backbone}
\end{figure}

\subsection{Definition and Architectures of Foundation Models}
A foundation model is a large-scale model pre-trained on broad and diverse data and subsequently adapted to multiple downstream tasks through prompting, fine-tuning, or lightweight adapters~\cite{Bommasani2021foundation}. Compared with traditional task-specific AI, the key shift is from ``train one model per task'' to ``pre-train once, and then adapt many times.'' In practice, foundation models are typically developed and deployed through large-scale pre-training, task-level adaptation, and deployment-time inference augmented by prompting, retrieval, or external tools. This paradigm improves data efficiency, transferability, and generalization ability.

In the wireless domain, a foundation model should not be defined solely by model size. Rather, it should learn reusable representations from large-scale wireless data, support adaptation across multiple tasks, scenarios, or system configurations, and exhibit non-trivial transferability beyond the distribution used for pre-training. Under this view, wireless foundation models include not only LLM-adapted wireless models, but also wireless-native models trained directly on CSI, CIR, I/Q streams, radio maps, interference graphs, or multi-modal sensing data. 

To support the massive scale and diverse data modalities required by this paradigm, the Transformer architecture has become the predominant backbone for foundation models. Its efficacy is primarily attributed to the self-attention mechanism, which enables the model to capture complex, long-range dependencies within data sequences. For an input token sequence matrix $\mathbf{X}_\text{in}\in\mathbb{R}^{n\times d_{\text{model}}}$, where $n$ is the sequence length and $d_{\text{model}}$ is the model dimension, the self-attention operation is formulated as
\begin{equation}
\mathbf{Q} = \mathbf{X}_\text{in}\mathbf{W}_Q,\quad \mathbf{K} = \mathbf{X}_\text{in}\mathbf{W}_K,\quad \mathbf{V} = \mathbf{X}_\text{in}\mathbf{W}_V,
\label{eq:qkv}
\end{equation}
\begin{equation}
\mathrm{Attention}(\mathbf{Q},\mathbf{K},\mathbf{V}) = \mathrm{softmax}\!\left(\frac{\mathbf{Q}\mathbf{K}^\top}{\sqrt{d_k}}\right)\mathbf{V},
\label{eq:self-attn}
\end{equation}
where $\mathbf{W}_Q\in\mathbb{R}^{d_{\text{model}}\times d_k}$, $\mathbf{W}_K\in\mathbb{R}^{d_{\text{model}}\times d_k}$, and $\mathbf{W}_V\in\mathbb{R}^{d_{\text{model}}\times d_v}$ are learnable projection matrices, yielding the query, key, and value matrices $\mathbf{Q},\mathbf{K}\in\mathbb{R}^{n\times d_k}$ and $\mathbf{V}\in\mathbb{R}^{n\times d_v}$, respectively. The scalar $d_k$ is the key dimension used for scaling, and $\mathrm{softmax}(\cdot)$ normalizes attention scores into attention weights.
Based on this shared backbone architecture, the development of foundation models has diverged into various families categorized by modality, as summarized in Table~\ref{tab:ai-fm-examples}.

In the text domain, many state-of-the-art LLMs adopt large-scale decoder-only generative architectures. Fig.~\ref{fig:transformer-backbone} highlights the key structural difference between an encoder-decoder Transformer and the decoder-only generative pre-trained Transformer (GPT)-style  design~\cite{vaswani2017attention, brown2020language}. In the computer vision domain, LVMs also adapt Transformer blocks to process image tokens. The representative Vision Transformer (ViT) pipeline tokenizes an image into patches and processes them using Transformer encoder blocks~\cite{dosovitskiy2021image}. Let an image of size $H \times W$ be partitioned into $N=HW/P^2$ patches of size $P \times P$. Then the input embedding of ViT can be formulated as
\begin{equation}
\mathbf{z}_0 = \left[ \mathbf{x}_{\mathrm{cls}}, \mathbf{x}_p^1\mathbf{E}, \mathbf{x}_p^2\mathbf{E}, \cdots, \mathbf{x}_p^N \mathbf{E} \right] + \mathbf{E}_{\mathrm{pos}},
\label{eq:vit-embed}
\end{equation}
where $\mathbf{x}_p^i$ is the $i$-th flattened image patch, $\mathbf{E}$ is the patch projection matrix, $\mathbf{x}_{\mathrm{cls}}$ is the learnable class token, and $\mathbf{E}_{\mathrm{pos}}$ denotes the positional embeddings.

Beyond single-modality backbones, multi-modal models facilitate the integration of information across diverse data sources. This family is primarily represented by vision-language models (VLMs), which are exemplified by the contrastive language-image pre-training (CLIP) \cite{radford2021learning} and large language and vision assistant (LLaVA)  \cite{li2025onevision} architectures. Unlike modality-specific designs, these models focus on the alignment and fusion of language and vision backbones. Such structural synergy enables the system to understand both modalities and perform generation tasks through the synchronization of linguistic and visual tokens. 


While Table~\ref{tab:ai-fm-examples} summarizes representative  foundation model families from the general AI perspective, wireless foundation models require a more system-oriented characterization. In particular, it is important to clarify what wireless data are used for pre-training, what pretext objectives are adopted, how adaptation is performed, and along which dimensions transferability is expected. Table~\ref{tab:wfm-compact-comparison} summarizes this wireless-domain perspective.

\begin{table*}[t]
	\caption{Compact Taxonomy of Wireless Foundation Model Classes}
	\label{tab:wfm-compact-comparison}
	\centering
	\footnotesize
	\setlength{\tabcolsep}{4.5pt}
	\renewcommand{\arraystretch}{1.15}
	\begin{tabular}{
			>{\raggedright\arraybackslash}p{0.14\textwidth}
			>{\raggedright\arraybackslash}p{0.15\textwidth}
			>{\raggedright\arraybackslash}p{0.15\textwidth}
			>{\raggedright\arraybackslash}p{0.15\textwidth}
			>{\raggedright\arraybackslash}p{0.15\textwidth}
			>{\raggedright\arraybackslash}p{0.15\textwidth}
		}
		\toprule
		\textbf{Model class} &
		\textbf{Pre-training data} &
		\textbf{Pretext objective} &
		\textbf{Adaptation method} &
		\textbf{Transfer dimension} &
		\textbf{Downstream tasks} \\
		\midrule
		
		LLM-adapted wireless FM &
		Text-pretrained LLMs with wireless tokens &
		Next-token prediction / task fine-tuning &
		LoRA / prompting / adapters &
		Task, scenario, channel condition &
		CSI, beam prediction, detection \\
		
		Wireless-native FM &
		CSI, CIR, I/Q streams, received waveforms, or spectrograms &
		Masked reconstruction / forecasting / contrastive learning &
		Fine-tuning / prompt conditioning &
		Task, antenna, bandwidth, environment &
		CSI, beam prediction, localization, sensing \\
		
		Graph/network FM &
		Interference graphs or network telemetry &
		Graph reconstruction / policy imitation &
		Prompting / fine-tuning / policy heads &
		Topology, density, traffic pattern &
		Power control, scheduling, slicing \\
		
		Agentic wireless FM &
		Standards, telemetry, tools, digital twins &
		Instruction tuning / RAG / tool use &
		Planning / tool invocation / multi-agent loops &
		Objective, policy, deployment &
		Orchestration, troubleshooting, network control \\
		
		\bottomrule
	\end{tabular}
\end{table*}

\subsection{Training Paradigms}
Despite differences in modality, modern foundation models share several common capabilities, including scalable representation learning, in-context generalization, cross-task transfer, and compositional reasoning. These capabilities emerge from a convergent training pipeline that integrates large-scale self-supervised pre-training, task-efficient adaptation, and inference-time augmentation.
The effectiveness of this pipeline relies on specific training objectives that are strategically coupled with each architecture and modality family. Under the generative learning paradigm, decoder-only LLMs are trained via autoregressive prediction to generate sequential content, while encoder-only language or vision models frequently employ masked-content reconstruction to learn robust representations. Meanwhile, multi-modal models utilize contrastive learning to achieve semantic alignment across different data samples or modalities. Training losses for these three learning paradigms can be respectively expressed as
\begin{equation}
\mathcal{L}_{\text{AR}} = - \sum_{t=1}^{T} \log p_{\theta}(x_t \mid x_{<t}),
\end{equation}

\begin{equation}
\mathcal{L}_{\text{mask}} = - \sum_{i \in \mathcal{M}} \log p_{\theta}(x_i \mid x_{\backslash \mathcal{M}}),
\end{equation}
\begin{equation}
\mathcal{L}_{\text{CL}} =
- \log
\frac{\exp(\mathrm{sim}(z_i, z_i^+)/\tau)}
{\sum_{j} \exp(\mathrm{sim}(z_i, z_j)/\tau)},
\end{equation}
where $\mathcal{L}_{\text{AR}}$, $\mathcal{L}_{\text{mask}}$, and $\mathcal{L}_{\text{CL}}$ denote autoregressive~\cite{brown2020language}, masked-modeling~\cite{devlin2019bert}, and contrastive objectives~\cite{oord2018representation}, respectively; $x_t$ is the token at step $t$; $x_{<t}$ is the preceding context; $p_{\theta}(\cdot)$ is the conditional probability parameterized by model parameters $\theta$; $\mathcal{M}$ is the masked index set; $x_{\backslash \mathcal{M}}$ denotes visible tokens; $z_i$ and $z_i^+$ are the anchor and its positive representation; $\mathrm{sim}(\cdot,\cdot)$ is a similarity function; and $\tau$ is the temperature hyper-parameter.

Building upon these pre-trained foundation models, recent advancements have introduced techniques to enhance reasoning and factual grounding during inference. For instance, chain-of-thought prompting \cite{wei2022chain} unlocks multi-step reasoning, while retrieval-augmented generation (RAG) \cite{lewis2020retrieval} mitigates hallucinations by grounding the model in external domain knowledge.
Ultimately, these capabilities are integrated into agentic frameworks (e.g., ReAct \cite{Yao2023react}), where the model transcends static prediction to perform interactive decision-making through tool invocation and environmental feedback.

\section{Foundation Models for Physical-Layer Processing} \label{sec:PHY}
While AI has significantly advanced physical-layer processing, current task-specific models still suffer from poor generalization and limited adaptability in dynamic wireless environments. Foundation models offer a promising pathway to address these limitations by learning transferable wireless representations and enabling efficient adaptation across tasks, scenarios, and system configurations. Against this background, this section explores the application of foundation models across three progressive paradigms. Section~\ref{subsec:Adapting Pre-trained} investigates adapting pre-trained foundation models to bypass manual NN design for specific signal processing tasks. Section~\ref{subsec:Wireless-Native Foundation} introduces wireless-native foundation models, which are pre-trained on wireless data to bridge modality gaps and capture universal physical representations. Section~\ref{subsec:Agentic-PHY} explores agentic foundation models, elevating the physical layer from static data processing to autonomous, reasoning-based orchestration. Finally, Section~\ref{subsec:datasets} reviews the cornerstone datasets driving this physical-layer intelligence.

\subsection{Adapting Pre-trained Foundation Models for Physical-Layer Processing} \label{subsec:Adapting Pre-trained}

\begin{table*}[htbp]
    \centering
    \caption{Summary of Adapting Pre-trained Foundation Models for Wireless Physical-Layer Tasks}
    \label{tab:phy_foundation_models}
    \renewcommand{\arraystretch}{1.3} 
    \begin{tabular*}{\textwidth}{@{\extracolsep{\fill}} l l c >{\raggedright\arraybackslash}m{2.5cm} >{\raggedright\arraybackslash}m{7.5cm} @{}}
        \toprule
        \textbf{Physical-Layer Task} & \textbf{Ref.} & \textbf{Year} & \textbf{Framework} & \textbf{Key Contribution} \\
        \midrule
        \multirow{4}{*}{CSI Estimation} & \cite{li2025llm4xce} & 2025 & LLM4XCE & Leverages feature-spatial attention and fine-tuning to recover spatial-channel representations. \\
         & 
         \cite{Guo2025lvm4csi} & 2025 & LVM4CSI & Applies LVMs to efficiently capture the spatial structures of CSI. \\
         & \cite{park2026sematic} & 2026 & Semantic Pilot & Utilizes LLM error correction capabilities for data-aided initial channel estimation. \\
        \midrule
        \multirow{3}{*}{CSI Feedback} & \cite{Cui2025llmCsiFeedback} & 2025 & PLM-CSI & Treats structured frequency-domain CSI tokens as corrupted words for denoising via pre-trained GPT-2. \\
         & \cite{Zhuang2025olamcf} & 2025 & OLAMCF & Utilizes a vision-oriented foundation model to refine random vector quantization codebooks. \\
        \midrule
        \multirow{7}{*}{Channel Prediction} & \cite{liu2024llm4cp} & 2024 & LLM4CP & Extracts propagation features via patch-based operations and CSI-specific attention. \\
         & \cite{li2026bridging} & 2025 & CSI-ALM & Bridges modality mismatch by mapping latent physical CSI features directly to word embeddings. \\
         & \cite{he2025scaLLM} & 2025 & SCA-LLM & Employs a 2D DCT-based multi-spectral adapter to combat Doppler shifts induced by mobility. \\
         & \cite{he2025sensing} & 2025 & Sensing-assisted LLM & Fuses multi-modal spatiotemporal features using ConvLSTM to ensure robust prediction. \\
        \midrule    
       \multirow{7}{*}{Beam Prediction} & \cite{Sheng2025beam} & 2024 & BP-LLM & 
        Pioneers explicit modality alignment by patch reprogramming to map historical beam-index/AoD sequences as text embeddings, leveraging LLMs' next-token prediction power. \\
         & \cite{lei2025llm} & 2025 & LLM-MM & Fuses heterogeneous multi-modal sensory inputs into a unified textual space for context-aware prediction. \\
         & \cite{liu2025large} & 2026 & CNN-GPT2 & Pre-trains LLM seamlessly with a CNN backbone to process unaligned raw physical signals directly. \\
         & \cite{kou2026beamvlm} & 2026 & BeamVLM & Reformulates prediction as an end-to-end generative semantic reasoning task via VLMs. \\
        \midrule        
        \multirow{5}{*}{Signal Detection} & \cite{yu2025large} & 2025 & Large sequence model & Integrates spatial geographic and AoA data to enhance zero-shot generalization across unseen users. \\
        & \cite{fan2025defined} & 2025 & DEFINED & Utilizes a decision feedback mechanism as pseudo-labels to refine in-context learning precision on-the-fly. \\
        & \cite{song2025turbo} & 2025 & Turbo-ICL & Introduces prompt augmentation to iteratively refine symbols using extrinsic information from channel decoders. \\
        \midrule
        \multirow{3}{*}{Multi-Task Processing} & 
        \cite{zheng2026large} & 2024 & Multi-task LLM & Unifies precoding, detection, and prediction within a single framework using LoRA fine-tuning. \\
        &
        \cite{ke2025signalllm} & 2025 & SignalLLM & Automates complex signal processing workflows via structured task decomposition, subtask planning, and execution. \\
 
        \bottomrule
    \end{tabular*}
\end{table*}

Adapting pre-trained foundation models to the physical layer introduces a fundamental challenge: bridging the inherent modality mismatch between continuous, dynamic physical signals and the discrete semantic space of these models. In this subsection, we explore how recent research overcomes this challenge across various physical-layer tasks, as summarized  in Table~\ref{tab:phy_foundation_models}. To provide a clear taxonomy, we first discuss the unified pipeline of CSI acquisition, including channel estimation, feedback, and prediction. Subsequently, we delve into transceiver design, e.g., beam prediction and signal detection, as well as emerging frameworks designed to handle multiple physical-layer tasks simultaneously.

\subsubsection{CSI Acquisition} 
Acquiring reliable CSI is a fundamental prerequisite for achieving the performance gains of massive MIMO systems. In practical network deployments, the CSI acquisition procedure is structurally partitioned into three critical phases: channel estimation, CSI feedback, and channel prediction. Recent literature individually customizes diverse pre-trained foundation models to overcome the distinct bottlenecks of each phase, including pilot overhead, limited feedback capacity, and severe channel aging.

First, for channel estimation, pre-trained foundation models are leveraged to improve accuracy and generalization under reduced pilot overhead. Specifically, to address the complexities of extremely large-scale MIMO, the LLM4XCE framework \cite{li2025llm4xce} leverages the semantic modeling capabilities of LLMs to recover spatial-channel representations. It introduces a parallel feature-spatial attention module to fuse the feature and spatial
dimensions of pilot signals. To bridge the modality gap between continuous wireless signals and the pre-trained LLM, LLM4XCE treats the spatial dimension (i.e., the number of antennas $M$) as the sequence length of tokens, and maps the extracted channel features to the hidden embedding dimension $d$. By adding learnable positional encoding, the continuous channel matrices are successfully transformed into a sequence of token embeddings. Finally, these tokens are processed by the LLM backbone through a lightweight fine-tuning strategy to achieve superior accuracy under hybrid-field conditions. 

While frameworks like LLM4XCE adapt LLMs to 1D sequences, angular-delay CSI inherently exhibits sparse, 2D image-like patterns. This structural characteristic motivates the application of LVMs for initial channel estimation, as demonstrated by LVM4CSI \cite{Guo2025lvm4csi}, which reveals that visual representations align more naturally with these spatial structures than language-based tokenization.
Moreover, multi-modal data, such as semantic information, can be integrated into LLM-based frameworks to further improve channel estimation performance. For example, a semantic pilot design for channel estimation is proposed in \cite{park2026sematic}. By leveraging the error correction capabilities of LLMs, reliable transmitted symbols are identified and subsequently used as additional pilots to facilitate channel estimation.

Second, the signaling overhead of CSI feedback is another critical bottleneck in practice. To achieve efficient compression and robust reconstruction, the PLM-CSI framework \cite{Cui2025llmCsiFeedback} leverages a pre-trained LLM by drawing an analogy between channel denoising and text correction. Specifically, it treats the coarsely reconstructed frequency-domain CSI tokens as ``corrupted words", enabling a pre-trained GPT-2 model to recover the channel by exploiting the strong correlation and sequential dependencies between adjacent subcarriers. Transitioning from this textual sequence modeling to a visual perspective, the OLAMCF framework \cite{Zhuang2025olamcf} utilizes a vision-oriented foundation model to refine conventional random vector quantization codebooks, underscoring that foundation model-driven feedback substantially benefits from the spatial representation capabilities of large visual backbones.

Finally, to combat channel aging in high-mobility scenarios, LLMs are deployed for temporal channel prediction. Specifically, given a sequence of $T$ historical channel observations $\{\mathbf{H}_{t}, \mathbf{H}_{t-1}, \dots, \mathbf{H}_{t-T+1}\}$, the goal is to forecast the channel states over the future $L$ time steps $\{\mathbf{H}_{t+1}, \dots, \mathbf{H}_{t+L}\}$, formally defined as
\begin{equation}
    \{\hat{\mathbf{H}}_{t+1}, \dots, \hat{\mathbf{H}}_{t+L}\} = f_{\text{pred}}(\mathbf{H}_{t}, \mathbf{H}_{t-1}, \dots, \mathbf{H}_{t-T+1}),
\end{equation}
where $f_{\text{pred}}(\cdot)$ denotes the LLM-based prediction function that maps historical channel observations to future channel estimates.
The fundamental challenge of bridging continuous wireless signals and the discrete text space is primarily resolved through customized feature alignment. For instance, the LLM4CP framework \cite{liu2024llm4cp} extracts propagation features via patch-based operations and CSI-specific attention, while CSI-ALM \cite{li2026bridging} maps latent CSI features directly to word embeddings. Beyond basic modality alignment, adapting LLMs to practical deployments requires tackling severe environmental dynamics. To handle Doppler shifts induced by mobility, the SCA-LLM architecture \cite{he2025scaLLM} employs a 2D discrete cosine transform (DCT)-based multi-spectral adapter to effectively preserve critical time-varying features. Furthermore, to combat severe fading and low signal-to-noise ratio (SNR) conditions, a sensing-assisted LLM method \cite{he2025sensing} within an ISAC architecture leverages ConvLSTM and cross-attention mechanisms to fuse spatiotemporal features from heterogeneous communication and sensing data, ensuring robust prediction even in harsh environments.

\begin{table}[t]
    \centering
    \caption{Performance comparison under matched and mismatched center frequency.}
    \label{tab:BP-LLM_performance}
    \renewcommand{\arraystretch}{1.3} 
    \begin{tabular}{llccccc}
        \toprule
         & & \multicolumn{5}{c}{\textbf{Prediction Time Step}} \\
        \cmidrule(lr){3-7}
        \textbf{Task} & \textbf{Model} & \textbf{1} & \textbf{3} & \textbf{5} & \textbf{7} &
        \textbf{9} \\
        \midrule
        \multirow{2}{*}{Matched} & BP-LLM & \textbf{0.950} & \textbf{0.900} & \textbf{0.865} & \textbf{0.857} & \textbf{0.805} \\
         & 
        ODE & 0.895 & 0.857 & 0.841 & 0.808 & 0.796 \\
        \multirow{2}{*}{Mismatched} & BP-LLM & \textbf{0.946} & \textbf{0.903} & \textbf{0.867} & \textbf{0.856} & \textbf{0.821} \\
         & 
        ODE & 0.783 & 0.744 & 0.700 & 0.608 & 0.579 \\
\bottomrule
\end{tabular}
\end{table}

\subsubsection{Beam Prediction}
In millimeter wave (mmWave) and emerging near-field communications, ultra-narrow beams overcome severe path loss but remain highly vulnerable to misalignment under high user mobility. Because dynamic beam tracking can be fundamentally formulated as a sequence forecasting problem, it naturally aligns with the robust sequence-modeling capabilities of LLMs. To bridge the modality mismatch between signals and textual tokens, recent studies employ three distinct paradigms: explicit modality alignment, domain-specific feature integration, and end-to-end generative reasoning.

\begin{figure*}[t]
    \centering
    \includegraphics[width=0.95\textwidth]{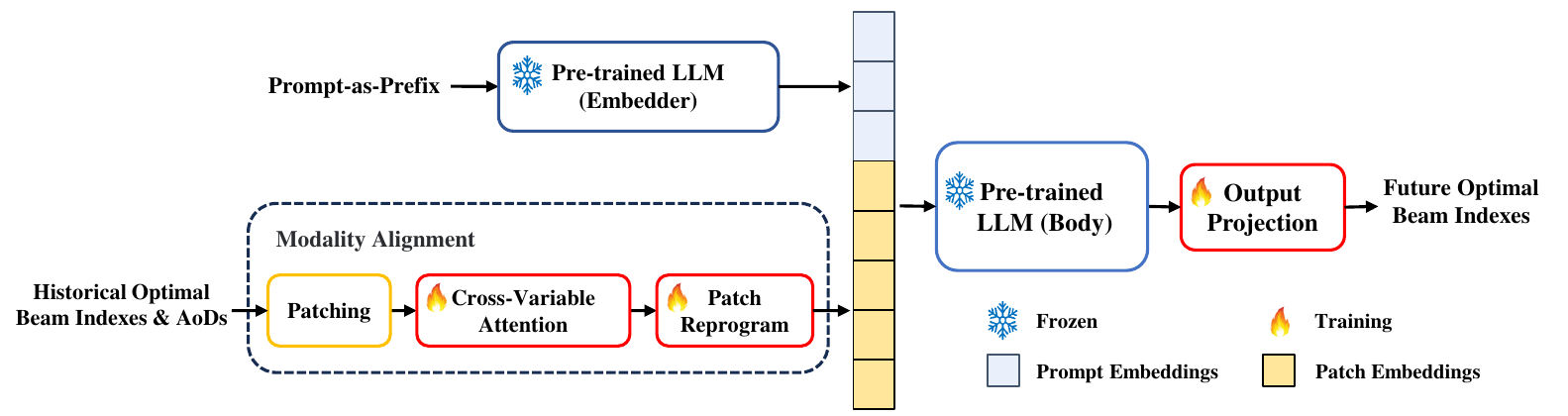}
    \caption{The architecture of BP-LLM \cite{Sheng2025beam}, which employs explicit modality alignment via patch reprogramming to map historical sequences of beam indices and AoDs into text embeddings.}
    \label{fig:beam_prediciotn}
\end{figure*}

The first paradigm tackles this mismatch through explicit modality alignment, transforming wireless data into an LLM-compatible semantic format. The landmark paper \cite{Sheng2025beam} proposes BP-LLM that utilizes patch reprogramming and batch vectorization to map physical radio frequency (RF) features, such as historical optimal beam indices and angles of departure (AoDs), directly into text embeddings, as illustrated in Fig.~\ref{fig:beam_prediciotn}. Coupled with a prompt-as-prefix strategy, BP-LLM effectively harnesses the LLM's inherent reasoning capabilities for temporal beam forecasting. As shown in Table~\ref{tab:BP-LLM_performance}, BP-LLM consistently outperforms existing learning-based methods under matched center frequency, validating the effectiveness of the proposed design. More strikingly, BP-LLM achieves remarkable generalization and robustness under mismatched center frequency. Furthermore, since optimal beam selection inherently depends on environmental geometry, this paradigm has expanded rapidly beyond RF modalities. For instance, 
LLM-MM \cite{lei2025llm} 
fuses heterogeneous sensory data from multi-modal environments into a unified textual space for context-aware prediction. Recent advancements in MLLM4BP \cite{mao2026beam} leverage locally available sensory data (e.g., LiDAR and camera) to facilitate the long-term prediction of beam indices, achieving performance comparable to that of perfect AoD-assisted beam prediction.

In contrast to explicit text mapping, the second paradigm pursues domain-specific feature integration. Rather than forcing physical signals to resemble language, CNN-GPT2 \cite{liu2025large} directly projects wireless features into the LLM's continuous embedding space, bypassing the intermediate step of text alignment. By utilizing a convolutional neural network (CNN) to process raw physical signals and executing customized pre-training directly on the LLM architecture, it bypasses token alignment entirely, enabling the model to capture non-linear dynamics and perform discriminative beam classification free from linguistic constraints.

Furthermore, the third paradigm radically reformulates beam prediction from a traditional classification problem into an end-to-end generative task. Leveraging VLMs, BeamVLM \cite{kou2026beamvlm} directly projects raw multi-modal sensory inputs into the model's latent space. Instead of merely outputting a predefined index, the model treats beam prediction as a multi-modal semantic reasoning process, autoregressively generating optimal beam configurations based on a holistic understanding of the dynamic environment.

\subsubsection{Signal Detection}
While traditional signal detection relies on accurate channel modeling and struggles with non-linear hardware impairments, adapting pre-trained foundation models offers an end-to-end, data-driven alternative. 
Fundamentally, LLM-based signal detection frames the problem as a sequence translation task. The received signals, along with the known pilots, are mapped into a unified prompt sequence. A prominent example is a recent study in \cite{yu2025large} that establishes a comprehensive framework for MIMO equalization by adapting a pre-trained GPT-2 architecture. Central to this framework is the concept of in-context learning. Unlike traditional machine learning that requires explicit parameter fine-tuning for new environments, in-context learning allows the model to infer task rules directly from the provided prompt sequence. Specifically, by observing the mappings between known pilots and received signals within the prompt, the pre-trained LLM implicitly learns the instantaneous channel characteristics. This enables the model to directly predict the subsequent transmitted discrete symbols without any parameter updates. By establishing a direct sequence-to-sequence mapping rather than relying on explicit channel inversion, this mechanism effectively avoids the amplification of channel estimation errors and noise enhancement typical of traditional receiver pipelines, while naturally absorbing complex non-linear hardware distortions.

Building upon this in-context learning paradigm, recent studies focus on dynamically augmenting the context to address practical system bottlenecks. For instance, since in-context learning heavily depends on sequence length, it is often restricted by limited pilot overhead. To mitigate pilot scarcity, the DEFINED framework \cite{fan2025defined} introduces a decision-feedback mechanism that appends previously detected symbols as pseudo-labels back into the prompt sequence. This dynamically extends the context to refine detection accuracy. Furthermore, achieving near-optimal performance in practical systems requires adapting to channel coding. The Turbo-ICL framework \cite{song2025turbo} bridges in-context learning and forward error correction using a soft-input soft-output equalizer. This framework augments prompts by converting decoder extrinsic information into additional contextual examples, enabling iterative symbol refinement across turbo iterations.

\subsubsection{Multi-Task Processing}
Since physical-layer processing involves various tasks, training and designing models for each of them separately requires high training resources, memory usage, and deployment costs. In response, pre-trained LLMs have been used for multi-task signal processing, which leverage their massive scale and unified architecture to translate diverse physical signals into a shared representation space. Hence, a single backbone model can capture common underlying patterns and dynamically adapt to various physical-layer objectives.
For instance, the work in \cite{zheng2026large} proposes a multi-task LLM for multi-user precoding, signal detection, and channel prediction, thereby unifying multiple operations within a single model. A fine-tuning strategy based on low-rank adaptation (LoRA) is introduced to reduce the memory requirements of
this model. Furthermore, a general-purpose and flexible LLM-based signal processing framework, known as SignalLLM, has been proposed in \cite{ke2025signalllm} to automatically handle complex signal processing tasks. This framework first performs signal processing task decomposition, signal processing subtask planning, and solution refinement. Subsequently, the subtasks are executed by either an LLM-assisted signal-processing reasoning module or an LLM-assisted signal-processing modeling module, depending on the specific subtask requirements. These solutions demonstrate strong performance across various complex signal processing tasks under extreme constraints, highlighting the potential of LLMs to automate complex signal processing workflows.

\subsection{Wireless-Native Foundation Models for Physical-Layer Processing} \label{subsec:Wireless-Native Foundation}
Despite the success of adapting general-purpose pre-trained foundation models for physical-layer tasks, as discussed in the previous subsection, the fundamental modality gap between discrete natural language tokens and continuous, multidimensional wireless signals makes the artificial alignment of these domains inherently suboptimal. 
Moreover, wireless signal processing is strictly bounded by microsecond-level latency requirements and limited computing resources, which are fundamentally incompatible with the massive parameter scale and text/video-centric execution of off-the-shelf LLMs/LVMs.

\begin{figure}[t]
    \centering
    \includegraphics[width=0.5\textwidth]{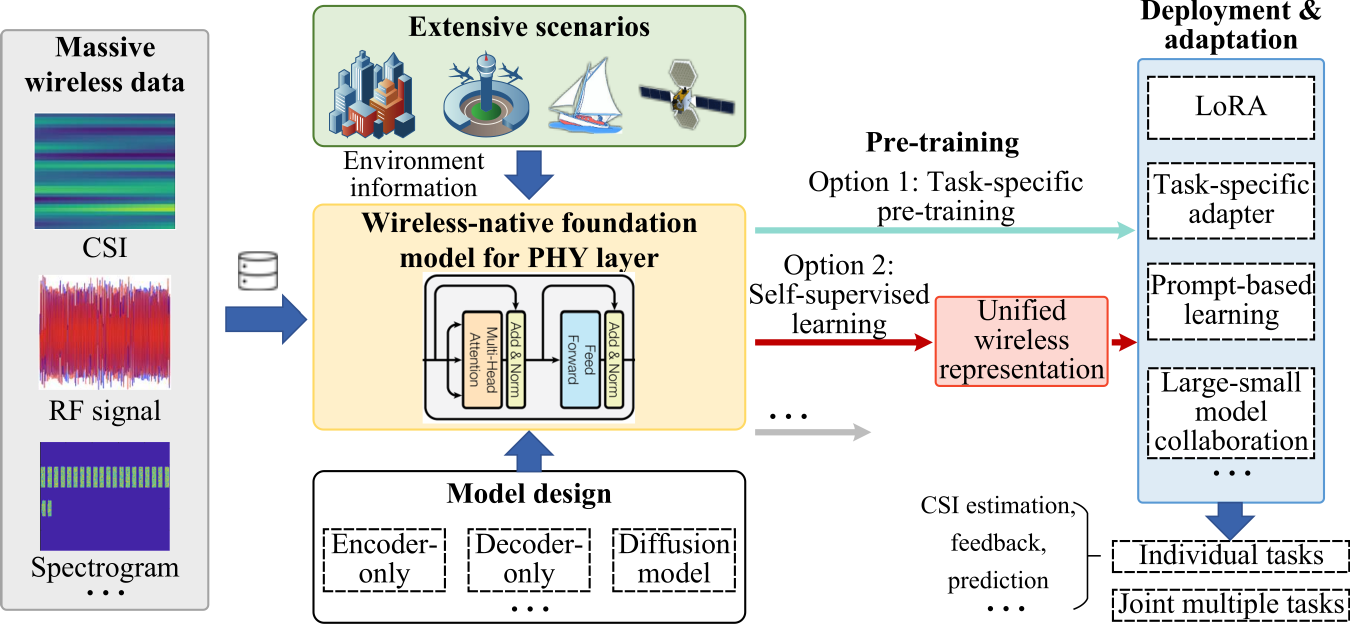}
    \caption{Framework of developing wireless-native foundation models for physical-layer processing. The core idea is to pre-train a foundation model from massive wireless data and diverse scenarios, enabling efficient adaptation to various downstream tasks.}
    \label{fig:wfm_phy}
\end{figure}

To resolve these architectural and physical bottlenecks, wireless-native foundation models for the physical layer are being developed, as demonstrated in Fig.~\ref{fig:wfm_phy}. By embedding physical priors directly into the neural architecture and pre-training on massive, domain-specific wireless data, this paradigm bypasses the semantic-to-signal modality mismatch. This enables the construction of highly compact, specialized foundation models that inherently understand the physical laws of electromagnetic propagation while satisfying stringent physical-layer constraints. Within this paradigm, CFMs represent a major class of wireless-native foundation models for channel understanding, aiming to learn transferable channel representations from diverse wireless observations containing channel information \cite{jiang2025towardscfm}.

\subsubsection{Task-Oriented Pre-trained Models}
Early wireless-native efforts primarily adopted task-oriented pre-training, where a model was pre-trained on large-scale wireless data but optimized for a particular category of physical-layer tasks. A representative work in this direction is WiFo \cite{liu2025wifo}. Initially designed as a space-time-frequency model for channel prediction, WiFo formulates channel forecasting as a unified reconstruction task.
Specifically, leveraging a MAE architecture, WiFo partitions the complex space-time-frequency channel tensor $\mathbf{H} \in \mathbb{C}^{T \times K \times N}$ into a set of 3D patches. By applying three masking strategies drawn from $\mathcal{M} \in \{\mathcal{M}_{\mathrm{random}},\mathcal{M}_{\mathrm{time}},\mathcal{M}_{\mathrm{frequency}}\}$, the model is forced to infer the hidden physical information using only the visible patches $\mathbf{H}_{\backslash \mathcal{M}}$. The core objective of its self-supervised pre-training is to minimize the reconstruction error over the masked regions $\mathbf{H}_{\mathcal{M}}$, which can be represented as
\begin{equation}
	\mathcal{L}=\mathbb{E}_{\mathbf{H},\mathcal{M}} \Big[
	\| \mathbf{H}_{\mathcal{M}} - f_{\theta}(\mathbf{H}_{\backslash \mathcal{M}})\|^2_F \Big],
\end{equation}
where $f_{\theta}(\cdot)$ denotes the WiFo encoder-decoder network, and $\|\cdot\|_F$ denotes the Frobenius norm.
Leveraging this mechanism, WiFo is pre-trained on massive CSI datasets and effectively captures the inherent 3D variations of wireless channels.
Consequently, WiFo demonstrates state-of-the-art zero-shot generalization across diverse antenna and system configurations without the need for immediate fine-tuning, providing a highly robust structural base for channel prediction tasks.

\begin{figure}[t]
\centering
\includegraphics[width=0.8\linewidth]{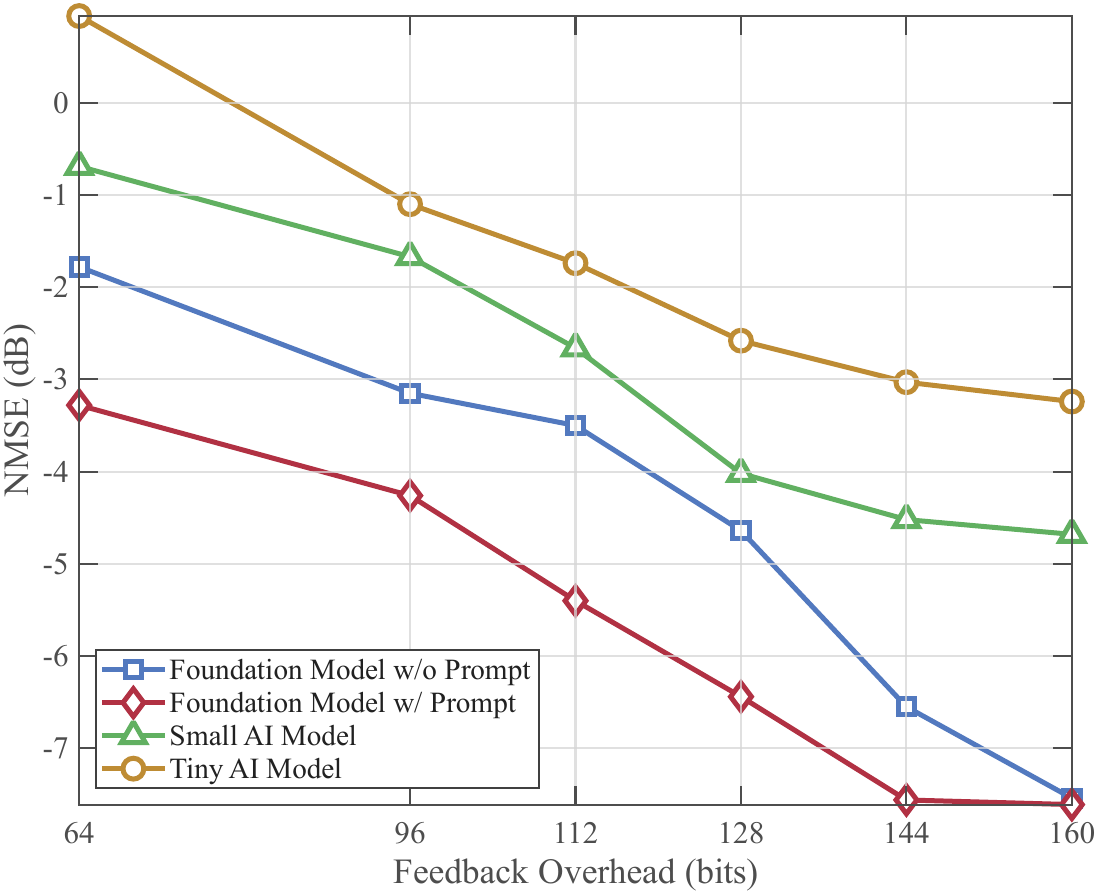}
\caption{NMSE versus feedback overhead for CSI feedback in unseen environments \cite{guo2025prompt}. The results highlight both the advantage of the foundation models over small and tiny models and the additional gain brought by prompt-guided reconstruction.}
\label{fig:prompt_csi_feedback}
\end{figure}

Beyond channel prediction, similar pre-trained approaches have also significantly advanced channel estimation \cite{zhou2025reducing,zhou2025generative} and CSI feedback \cite{guo2025prompt,liu2026wifocf}.
For channel estimation, task-oriented pre-trained models have been specifically customized to tackle the critical challenges of substantial pilot overhead and limited cross-scenario generalization. By pre-training on large-scale wireless channel data, these models capture universal channel representations, such as the predictive priors in \cite{zhou2025reducing} and the deep generative priors extracted by diffusion models in \cite{zhou2025generative}. These robust structural priors enable high-fidelity channel reconstruction even from extremely sparse or noisy pilot observations, significantly reducing pilot overhead while maintaining strong adaptability across unseen environments.
In the context of CSI feedback, the prompt-enabled large AI model in \cite{guo2025prompt} demonstrates that it is beneficial to inject environment-specific information into the reconstruction process. Specifically, the decoder takes as a prompt the mean channel magnitude in the angular-delay domain of the current environment, so that CSI reconstruction is guided not only by the received feedback bits but also by slowly varying environmental statistics. This prompt mechanism improves feedback accuracy with only a very small signaling overhead. As shown in Fig.~\ref{fig:prompt_csi_feedback}, the pre-trained model provides clear gains over smaller and tiny models, while the proposed prompt mechanism further reduces normalized mean-squared error (NMSE) by injecting environment-specific knowledge.
A related development is WiFo-CF \cite{liu2026wifocf}, which pushes the model design further toward a unified CSI feedback foundation model. Instead of tailoring one network to one fixed configuration, WiFo-CF is designed to handle heterogeneous feedback dimensions, rates, and channel distributions within a single pre-trained framework, overcoming the limitations of configuration-specific models. 
Along this line of thought, WiCC-Net \cite{jing2026signal} proposes a Transformer-based wireless foundation model tailored for CSI compression, which was pre-trained from scratch on over 172 million synthetic CSI samples across diverse scenarios using a self-supervised masked reconstruction objective. It adopts an encoder-only architecture equipped with a dedicated self-attention compressor module to produce highly compact yet informative representations.

Beyond CSI processing, task-oriented pre-training has expanded to other physical-layer applications. In signal recognition and spectrum management, models pre-trained directly on raw I/Q streams or baseband spectrograms, such as self-supervised radio models \cite{aboulfotouh2024self} and SpectrumFM \cite{zhou2025spectrumfm}, have demonstrated strong generalization across modulation classification, spectrum sensing, and anomaly detection tasks.
Concurrently, in the realm of wireless sensing and localization, pre-trained models are being developed to extract robust spatial and geometric information from massive unlabeled data \cite{salihu2024self}. By alleviating the reliance on site-specific calibration and extensive labeled datasets, these models demonstrate remarkable cross-scenario generalization. 
Furthermore, this pre-training paradigm has been applied to core transceiver operations, with dedicated models emerging for heterogeneous multiuser demodulation \cite{yang2026wifo} and massive MIMO precoding \cite{emery2025foundation}, demonstrating its potential across the physical-layer pipeline.

Nevertheless, deploying these pre-trained models in practical systems requires addressing performance degradation caused by imperfect CSI. To tackle this issue, the Filter-and-Attend paradigm \cite{wang2025filter} introduces a robustness-oriented design. Rather than directly processing noisy inputs, it first suppresses noise-plus-interference components in the received signals via projection matrices constructed from coarse CSI estimates, and subsequently performs attention-based CSI completion for clean channel feature extraction. This approach achieves consistent performance improvements across multiple downstream tasks, including channel prediction and localization.
These studies show that task-oriented pre-trained models are becoming less tied to a single system configuration and increasingly capable of exploiting shared propagation, signal, and environmental structures. As their learned representations become transferable across multiple task categories, these models move toward the universal, task-agnostic foundation models discussed next.

\subsubsection{Universal Task-Agnostic Models}
Despite their zero-shot capabilities within a specific task category, task-oriented pre-trained models are generally tailored to individual communication modules, and their pre-training objectives and structural boundaries limit broader cross-task applicability.
To truly unlock the potential of foundation models, the physical layer demands universal models that can be applied across a wide range of tasks/modules via few-shot or even zero-shot learning. The core philosophy shifts towards extracting \textit{unified, task-agnostic physical-layer representations} from massive volumes of unlabeled wireless data via \textit{self-supervised learning}, alleviating the need for substantial task-specific labeled data. 
These inputs may include explicit CSI or CIR as well as implicit observations such as I/Q samples and received waveforms. 
By employing methodologies such as masked modeling, next-token prediction, and contrastive learning, these models, including CFMs \cite{jiang2025towardscfm}, project raw wireless signals into a rich, contextualized latent space, which can subsequently drive a vast array of downstream physical-layer tasks.

\begin{figure*}
\centering{\includegraphics[width=0.9\textwidth]{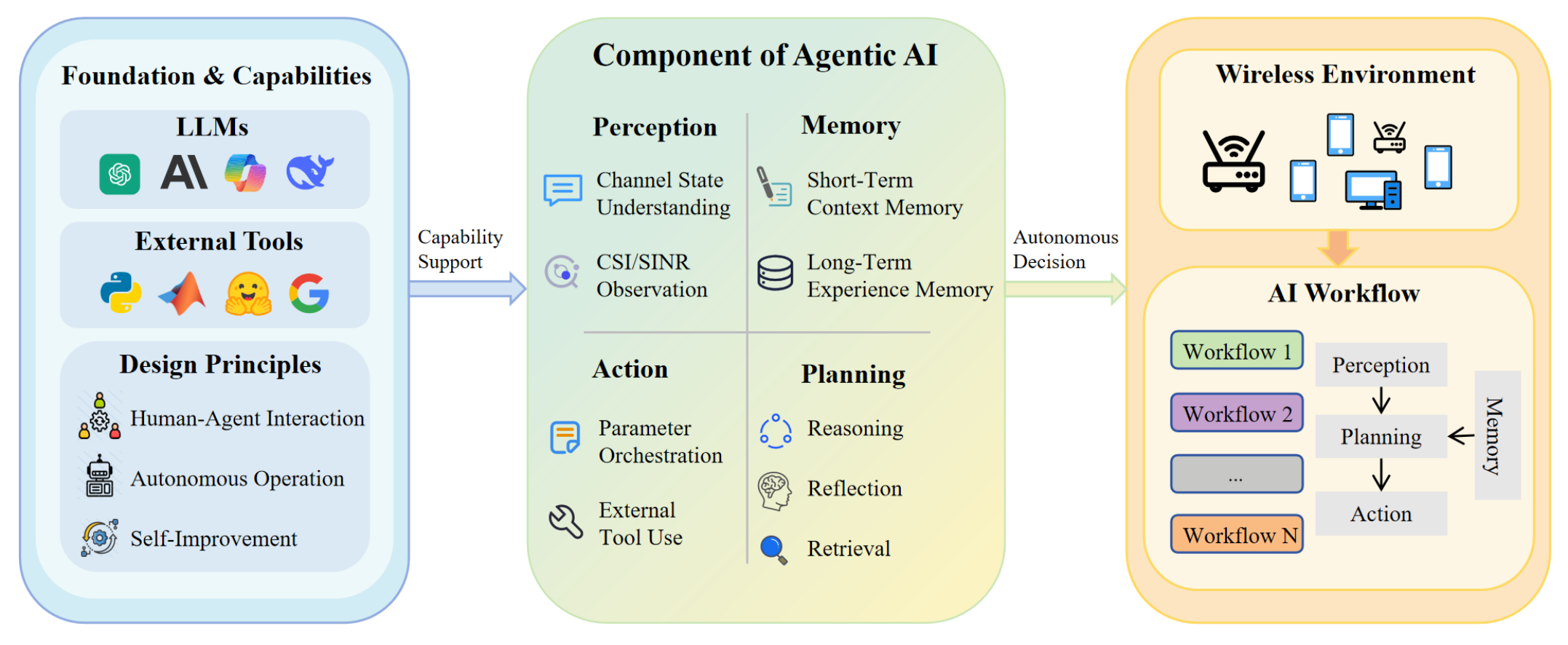}}
\caption{The architecture of agentic AI in physical-layer processing, adapted from \cite{Tong2025wirelessagent}. The framework is composed of foundational capabilities, the agentic AI component, and wireless-environment-oriented AI workflows. The foundational capabilities provide the supporting basis for agents, while the agentic AI component leverages perception, memory, planning, and action to understand wireless states, manage historical experiences, reason over task objectives, and orchestrate transmission parameters. The AI workflow organizes task execution and decision-making procedures, enabling coordinated agent operations for autonomous physical-layer optimization.}
\label{fig_agent}
\end{figure*}

The large wireless model (LWM) proposed in \cite{alikhani2024large} stands as a landmark in this universal paradigm. Conceived as a task-agnostic feature extractor, LWM is pre-trained on massive unlabeled wireless datasets using masked channel modeling. Specifically, the pre-training objective minimizes the mean-squared error (MSE) between the original channel patches and their reconstructions, formulated as
\begin{equation}
\mathcal{L}_{\mathrm{MCM}} = \frac{1}{|\mathcal{P}|} \sum_{i \in \mathcal{P}} \left\| \mathbf{W}_{\mathrm{dec}} \mathbf{e}_{\mathrm{LWM},i} - \mathbf{p}_i \right\|^2,
\end{equation}
where $\mathcal{P}$ represents the set of masked patches, $\mathbf{e}_{\mathrm{LWM},i}$ denotes the LWM's high-dimensional embedding of the $i$-th masked patch, $\mathbf{W}_{\mathrm{dec}}$ denotes the weight matrix of a linear decoder mapping the embedding back to its original size, and $\mathbf{p}_i$ is the original patch value.
By forcing the network to reconstruct randomly masked channel patches, LWM learns the intricate, underlying physical correlations, such as spatial multipath geometries, AoA/AoD clusters, and wideband spectral dependencies. 
Rather than outputting specific task labels, this model generates rich, contextualized embeddings that can be integrated into diverse downstream wireless tasks, thereby enhancing performance with remarkable data efficiency, particularly in label-limited scenarios.

Building upon universal representations, wireless-native foundation models have further evolved along two complementary directions: transferable channel representation learning and cross-functional integration. Representative CFMs demonstrate that transferable representations can be learned from different forms of channel information. CSI-CLIP \cite{jiang2025mimo} exploits the physical consistency between CIR and CSI views of the same propagation process to facilitate transfer to beam prediction, positioning, and LoS/NLoS identification. ContraWiMAE \cite{guler2025multi} combines masked reconstruction with representation alignment to support multiple channel-related tasks. Although their pre-training designs differ, these models share the goal of transferring learned channel representations beyond a single task or system configuration. Beyond such channel-centric transfer, wireless-native foundation models have also expanded toward joint multi-task execution, cross-functional integration, and mixed-timescale data processing. For instance, generative pre-trained multi-task learning frameworks like WirelessGPT \cite{yang2025wirelessgpt} seamlessly accommodate ISAC systems since the constructed task-agnostic representation unifies communication and sensing functionalities, thereby supporting highly divergent tasks, ranging from signal detection to dynamic target tracking and environment reconstruction.
The recently proposed in-context wireless large model (ICWLM) \cite{wen2026icwlm} continues this trend towards universal physical-layer intelligence by developing a wireless-native causal Transformer backbone that uniquely leverages in-context learning, jointly solving diverse tasks like multi-user precoding and channel prediction, and demonstrating exceptional adaptation to unseen system configurations via minimal demonstration pairs rather than extensive fine-tuning.
This line of research has been further extended to develop a versatile multi-task predictor in wireless systems~\cite{sheng2025wireless}. Recognizing the diverse temporal dynamics of different prediction variables, ranging from rapidly fluctuating CSI to slowly varying network traffic, this unified predictive foundation model incorporates univariate decomposition techniques to standardize heterogeneous tasks within a single framework. Furthermore, by employing granularity encoding for interval awareness and a causal Transformer backbone with a flexible patch masking strategy, the model seamlessly processes arbitrary input history lengths and diverse time granularities.

\subsubsection{Deployment and Adaptation Strategies}
With the superiority of universal wireless-native foundation models such as CFMs, a critical research direction is the \textit{deployment strategy} in the physical layer: \textit{how to transfer massive wireless knowledge to specific tasks or environments under stringent deployment constraints?} Given that full-parameter fine-tuning is computationally prohibitive, adapting parameter-efficient fine-tuning (PEFT) techniques, such as LoRA \cite{liu2025llm4wm} and downstream-task-specific adapters, has become an important pathway. Furthermore, prompt-based learning \cite{guo2025prompt} is being actively explored to guide universal models to adapt to specific tasks without affecting model weights, thereby enabling fast few-shot or even zero-shot adaptation. 

In addition, large and small model collaboration \cite{cui2025large} has emerged as an effective framework to achieve highly efficient environment-specific adaptation. This framework exploits the pre-trained wireless-native foundation model as a frozen, universal channel knowledge base, while small AI models are employed as lightweight plugins. For instance, the pre-trained large model can provide a generalized initial signal reconstruction (e.g., for CSI feedback), and the small AI models are explicitly trained to capture local, environment-induced physical shifts. By dynamically mapping these learned shifts back to correct the large model's outputs, this collaborative architecture achieves remarkable environment-specific performance gains, enabling ultra-fast adaptation speed and significantly lowering data collection requirements.

\subsection{Agentic Foundation Models for Physical-Layer Automation}\label{subsec:Agentic-PHY}
While the adapted and wireless-native foundation models discussed in previous subsections have demonstrated remarkable capabilities in modeling complex wireless signals, they predominantly function as advanced feature encoders. Operating essentially as ``black boxes," these models focus on extracting static, latent representations for specific numerical optimizations (e.g., channel prediction), yet they inherently lack proactive intelligence, transparency, and dynamic adaptability. However, the emerging 6G paradigm requires fully autonomous and intent-driven networks capable of seamlessly adapting to highly heterogeneous and unpredictable environments. To meet these demands, agentic foundation models are being developed to elevate the physical layer into an autonomous system. Rather than merely executing static mathematical mappings, wireless LLM agents can actively perceive dynamic channel environments, logically reason through multi-objective task requirements, and autonomously orchestrate transmission parameters, all while providing crucial interpretability for their actions \cite{Tong2025wirelessagent}, as illustrated in Fig.~\ref{fig_agent}.

As an example, conventional beam management in mmWave MIMO systems heavily relies on opaque numerical regression or overhead-intensive beam sweeping while agent-driven approaches leverage the semantic reasoning of LLMs to achieve explainable beam tracking. For instance, BeamAgent \cite{beamagent} proposes a closed-loop framework that integrates requirement interpretation, problem formulation, numerical tool invocation, and feedback-based refinement. By decoupling semantic reasoning from numerical optimization, this framework enables the agentic foundation model to focus on task understanding and evaluation, while assigning the concrete optimization process to external tools. By making the decision-making process transparent and physically traceable, these agentic frameworks directly resolve the unreliability of traditional black-box machine learning in dynamic wireless environments. 

Beyond single-task reasoning, agentic foundation models can also serve as system-level orchestrators that break the traditional silos of individual physical-layer modules \cite{llm_6g_orchestrator}. In this setting, LLM-enabled agents act as intent-aware cognitive controllers that combine raw CSI with high-level semantic intents to jointly optimize multiple transmission components. For example, AgenCom autonomously evaluates end-to-end feedback to coordinate modulation, precoding, and equalization \cite{li2026agentic}. This cross-module coordination enables the agent to dynamically tailor link configurations according to evolving reliability or throughput objectives, thereby extending physical-layer operation from isolated decisions to coordinated pipelines. 


To ensure these intent-driven physical-layer optimizations are grounded in realistic network states, digital twin technologies are increasingly integrated with agentic LLMs~\cite{haider2026llm}. By establishing a bidirectional feedback loop between the physical environment and its virtual replicas, frameworks such as LLM-DTNet enable the LLM to translate textual policies into explicit optimization variables. At the foundational physical layer, this allows the agentic system to dynamically refine 3D ray-tracing channel models through site-specific radio material recommendations and autonomously execute context-optimal precoding schemes, which directly translates to measurable gains in spectral efficiency and bit error rates. Furthermore, architectures such as AgentRAN utilize a hierarchy of LLM-powered agents to translate natural language intents into coordinated control loops, directly propagating high-level network objectives down to the physical layer for precise physical-layer parameter tuning \cite{elkael2026agentran}. Through continuous environment interaction and structured strategy negotiation, these intent-driven frameworks lay the foundation for fully autonomous, self-optimizing 6G networks.

\subsection{Datasets} \label{subsec:datasets}

The remarkable capabilities of physical-layer foundation models critically depend on massive, high-quality data. To fuel both the pre-training and fine-tuning of wireless foundation models, the community has contributed a diverse array of datasets, which generally fall into two complementary paradigms: simulation-based environments and real-world measurements.
Simulation-based datasets are widely adopted because they easily leverage advanced ray-tracing techniques to generate the full-dimensional CSI required by foundation models at scale. Early frameworks like ViWi \cite{viwi} and DeepMIMO \cite{deepmimo} established the baseline by offering generic pipelines for synthetic channel generation. Recent advancements have evolved toward high-fidelity digital twins and massive multi-modal integration. For example, SynthSoM \cite{synthsom} simulates complex air-ground cooperations, and Multimodal-Wireless \cite{mmw} establishes a large-scale dataset for multi-modal sensing and communication. Furthermore, the recently proposed LH-CSI dataset \cite{liu2025foundation} provides an unprecedented 11.6 billion heterogeneous 3D CSI data points tailored to support versatile multi-task foundation models.

\begin{table*}[t]
    \centering
    \caption{Summary of Representative Datasets for Wireless Foundation Models}
    \label{tab:datasets_comparison}
    \renewcommand{\arraystretch}{1.3} 
    \begin{tabular*}{\textwidth}{@{\extracolsep{\fill}} l l c c >{\raggedright\arraybackslash}m{2.55cm} >{\raggedright\arraybackslash}m{3.1cm} >{\raggedright\arraybackslash}m{4.85cm} @{}}
        \toprule
        \textbf{Type} & \textbf{Dataset} & \textbf{Year} & \textbf{CSI} & \textbf{Sensor Modalities} & \textbf{Scenario / Topology} & \textbf{Key Feature / Application} \\
        \midrule
        \multirow{7}{*}{Simulation} & ViWi \cite{viwi} & 2019 & Yes & RGB, Depth, LiDAR & Indoor / Outdoor & Aligned synthetic views \& ray-tracing \\
         & DeepMIMO \cite{deepmimo} & 2019 & Yes & None & Generic topologies & Parameterized RF massive MIMO \\
         & SynthSoM \cite{synthsom} & 2025 & Yes & RGB, Depth, LiDAR, Radar & Air-ground & Machine synesthesia, channel fading \\
         & Multimodal-Wireless \cite{mmw} & 2025 & Yes & RGB, Depth, LiDAR, Radar, Inertial Measurement Unit & High-mobility\;environment & Fully synchronized CSI and multi-modal data \\
         & LH-CSI \cite{liu2025foundation} & 2025 & Yes & None & Heterogeneous\;configurations & 11.6B\;CSI\;points\;across\;multiple datasets \\
        \midrule
        \multirow{3}{*}{Real-world} 
         & E-flash \cite{e-flash} & 2022 & No & RGB, LiDAR & Dynamic urban canyon & mmWave MIMO beam selection \\
         & DeepSense 6G \cite{DeepSense} & 2022 & No & RGB, LiDAR, Radar & Diverse deployments & Large-scale synced multi-modal \\
         & CSI-bench \cite{csibench} & 2025 & Yes & None & Diverse indoor & Multi-task Wi-Fi sensing benchmark \\
        \bottomrule
    \end{tabular*}
\end{table*}

While simulations offer high-fidelity and easily accessible channels, real-world datasets remain indispensable for capturing unmodeled anomalies, hardware impairments, and genuine environmental dynamics. For instance, real-world datasets, such as e-flash \cite{e-flash} and the large-scale DeepSense 6G \cite{DeepSense}, provide invaluable physical authenticity. However, due to real-world constraints in practical setups, they typically yield empirical metrics, such as received signal strength, optimal beam indices, or raw waveforms, rather than explicitly decoupled CSI. Bridging this gap, recent efforts like CSI-Bench \cite{csibench} leverage commercial edge devices to capture over 460 hours of real-world, in-the-wild Wi-Fi CSI, providing a highly generalized benchmark for multi-task human sensing. 
A detailed comparison of these datasets, highlighting their modalities, scenarios, and CSI availability, is provided in Table~\ref{tab:datasets_comparison}.

\section{Foundation Models for Wireless Resource Allocation} \label{sec:RA}

Unlike physical-layer processing that usually maximizes individual link efficiency using localized CSI, wireless resource allocation targets dense multi-user environments. This fundamental shift necessitates reasoning over mutually coupled interference topologies rather than isolated channels, alongside balancing system-level objectives, such as proportional fairness and quality of service (QoS) guarantees, which are largely absent in single-user designs. To address these network-level challenges, this section explores the application of foundation models across three paradigms. Section~\ref{subsec:pre-trained-FM4RA} investigates adapting pre-trained foundation models for interference management. To overcome the inherent cross-domain modality gap, Section~\ref{subsec:Wirless-FM4RA} introduces wireless-native foundation models, explicitly pre-trained on domain-specific radio graphs to capture universal interference representations. Finally, Section~\ref{subsec:Agentic-FM4RA} explores agentic foundation models, extending the paradigm from static solvers to autonomous, closed-loop resource orchestration frameworks driven by multi-agent collaboration and user intents.

\subsection{Adapting Pre-trained Foundation Models for Resource Allocation}\label{subsec:pre-trained-FM4RA}

\begin{figure}[t]
    \centering
    \includegraphics[width=0.5\textwidth]{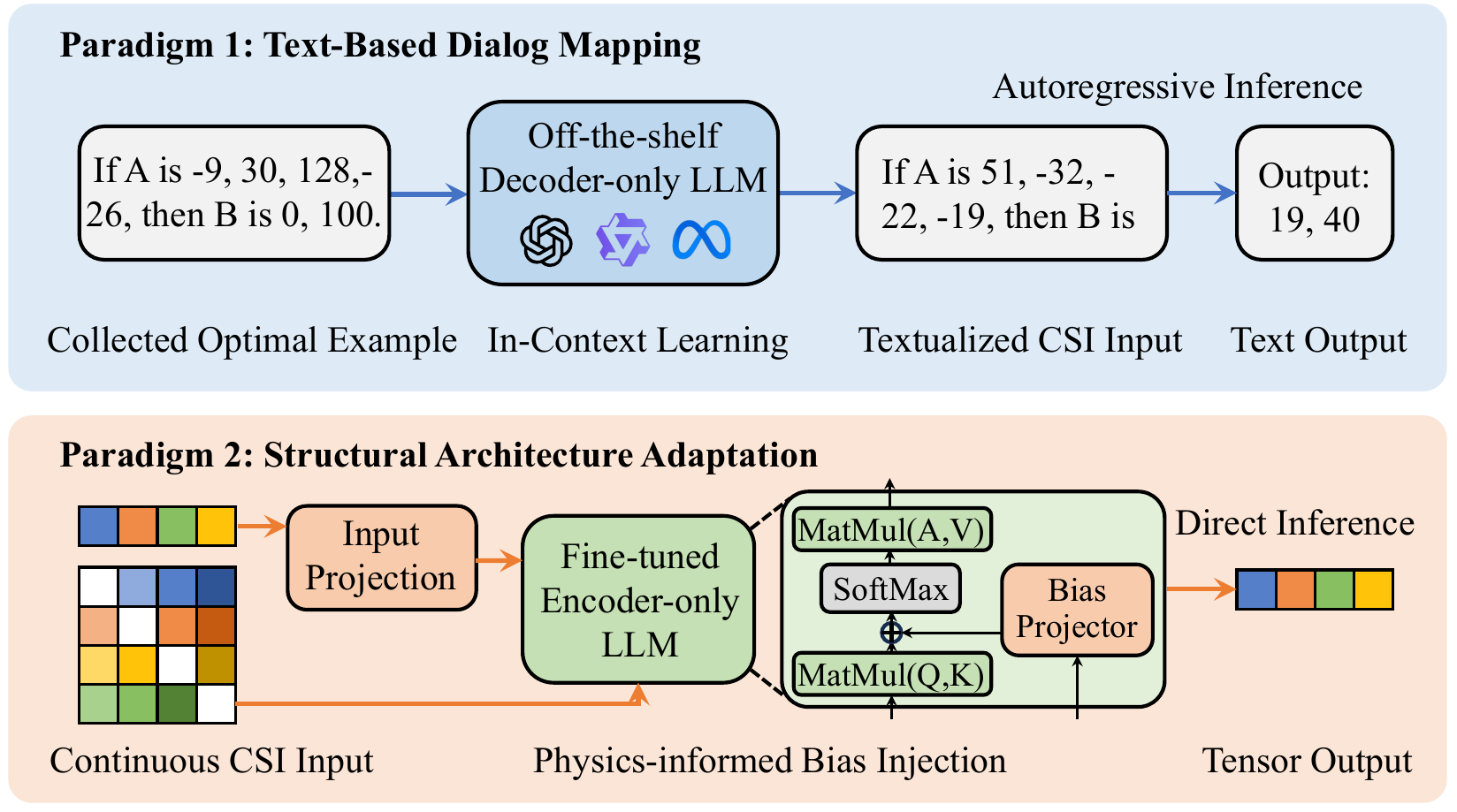}
    \caption{Paradigms of foundation model adaptation for wireless resource allocation: (top) text-based dialogue mapping utilizing off-the-shelf LLMs \cite{lee2026llm}, and (bottom) structural architecture adaptation via physics-informed backbones \cite{wang2026wireless}.}
    \label{fig:LLM4RA_comparation}
\end{figure}

Wireless resource allocation involves judiciously allocating limited communication resources, such as transmission power, spectrum, and time slots, among multiple users, where the core challenge lies in managing the intricate multi-user interference within a dense environment. 
For instance, the power control problem is cast as maximizing a network-wide utility function $\beta(\cdot)$ (e.g., sum-rate or proportional fairness) subject to individual power constraints $p_k \in [0, P_{\max}]$, expressed as
\begin{equation} \label{eq:resource_allocation_objective}
    \max_{\mathbf{p}} \sum_{k=1}^K \beta \left( \log_2 \Big(1 + \frac{p_k |h_{kk}|^2}{\sum_{j \neq k} p_j |h_{kj}|^2 + \sigma^2}\Big) \right),
\end{equation}
where $|h_{kk}|^2$ and $|h_{kj}|^2$ denote the direct and cross-link channel gains, respectively, and $\sigma^2$ is the noise power. The mutually coupled interference among users, as captured in the denominator in \eqref{eq:resource_allocation_objective}, shares similarities with the contextual dependencies among different words in natural language. Both of them demand complex relational reasoning, and therefore the multi-head self-attention mechanisms of Transformer architectures, widely used in natural language modeling, can be employed to disentangle these multi-user interactions.To this end, adapting pre-trained foundation models to address wireless resource allocation has followed two paradigms, as illustrated in Fig.~\ref{fig:LLM4RA_comparation}. Early approaches formulate physical optimization as a text-based dialogue task, whereas recent advancements structurally adapt the underlying neural architectures to directly process continuous physical variables.

The first paradigm treats resource optimization as sequence-to-sequence text mapping using off-the-shelf, decoder-only generative models. By employing few-shot in-context learning, these models perform static reasoning without requiring LLM model parameter updates. For example, formatting sample pairs of channel states and desired transmit powers as text prompts allows the model to contextualize the problem and reason over the non-linear trade-offs between power allocation and link performance. When given new CSI inputs, the LLM model generates the corresponding resource allocation results, often yielding near-optimal performance~\cite{lee2026llm}. Furthermore, the LLM-RAO framework \cite{noh2025adaptive} introduces a closed-loop approach where the language model generates a heuristic initial solution that an external solver subsequently evaluates. Feeding these quantitative scores back as dialogue prompts enables iterative self-refinement, effectively compensating for the shortcomings of pure text generation.

\begin{figure}[t]
    \centering
    \includegraphics[width=0.85\columnwidth]{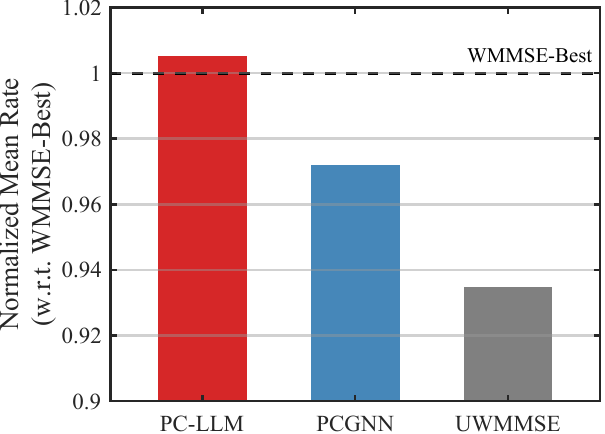}
    \caption{Max-sum-rate performance of adapting a pre-trained LLM to wireless power control. Bars show the averaged normalized mean rate, with WMMSE-Best shown as the dashed reference line.}
    \label{fig:pcllm_ra_results}
\end{figure}

Nevertheless, text-based adaptations still suffer from autoregressive inference latency, and it is conceptually difficult to appreciate and confirm that the pre-trained language model truly understands the wireless resource allocation problem and always generates the right decision outputs. To address these limitations, PC-LLM~\cite{wang2026wireless} shifts to an encoder-only backbone and produces power-allocation decisions in a single forward pass. Instead of using standard text tokenization, it injects the channel gain matrix into the self-attention scores through an interference-aware bias projector, thereby turning the pre-trained language backbone into a topology-aware graph reasoning engine. Fig.~\ref{fig:pcllm_ra_results} provides a representative max-sum-rate comparison among PC-LLM, WMMSE~\cite{christensen2008wmmse}, PCGNN~\cite{shen2023graph}, and unfolding WMMSE (UWMMSE)~\cite{chowdhury2021unfolding}. WMMSE is evaluated over 100 independent random initializations and WMMSE-Best denotes the best result among these 100 runs. The results show that PC-LLM matches or slightly exceeds the WMMSE-Best reference and clearly outperforms the state-of-the-art baseline PCGNN, highlighting the effectiveness of adapting pre-trained foundation models to interference-aware resource allocation.

While this structural adaptation confirms that existing pre-trained models can support physical interference management, their weights are still inherited from models pre-trained on linguistic corpora. This modality mismatch motivates a shift from cross-domain adaptation to wireless-native foundation models, which are pre-trained directly on massive wireless data and discussed in the following subsection.

\begin{table*}[t]
\centering
\begin{threeparttable}
\caption{Summary of Agentic Foundation Models for Wireless Resource Allocation}
\label{tab:agentic_fm_resource_allocation}
\renewcommand{\arraystretch}{1.2}
\setlength{\tabcolsep}{3pt}

\begin{tabular}{ccc c >{\raggedright\arraybackslash}p{2.5cm} c c >{\raggedright\arraybackslash}p{4.6cm}}
\toprule
Reference & Year & Objective & Method & Task & Arch. & Opt. Mode & Key Contribution \\
\midrule

\cite{lotfi2025orangouide} 
& 2025 
& Metric-Driven 
& RAG-Augmented RL 
& Network slicing and scheduling 
& Cen. 
& RAG+RL 
& Uses RAG for semantic state representation in QoS-aware slicing. \\

\cite{lee2025aidriven} 
& 2025 
& Metric-Driven 
& Distributed MA 
& Power control 
& Hor. 
& LLM agent 
& Achieves CSI-free distributed power control via inter-agent experience sharing. \\

\cite{Fan2025mapc} 
& 2025 
& Metric-Driven 
& Dialogue-Based MA 
& Transmission scheduling 
& Hor. 
& LLM agent 
& Negotiates Wi-Fi AP schedules via natural-language dialogue. \\

\cite{zheng2026rl_opt} 
& 2026 
& Metric-Driven 
& LLM-Empowered RL 
& Bandwidth, power, and computing resource allocation 
& Cen. 
& LLM+RL 
& Enhances RL feature abstraction, reward design, and policy interpretation. \\

\cite{bao2026llm} 
& 2026 
& Metric-Driven 
& Hierarchical LLM-RL 
& Power allocation and scheduling 
& Vert. 
& LLM+RL 
& Enables cross-timescale control via Non-RT LLMs and Near-RT RL. \\

\midrule

\cite{han2026multillm} 
& 2026 
& Intent-Driven 
& Role-Specialized MA 
& Slicing and bandwidth allocation 
& Cen. 
& LLM agent 
& Allocates multi-BS bandwidth interactively using role-specialized LLM agents. \\

\cite{li2026comagent} 
& 2026 
& Intent-Driven 
& Workflow-Driven MA 
& Beamforming 
& Vert. 
& LLM+Opt. 
& Autonomously formulates, codes, and validates beamforming problems from parsed intents. \\

\cite{liu2026lameta} 
& 2026 
& Intent-Driven 
& Intent-Distilled RL 
& Computational resource routing 
& Vert. 
& Distill.+RL 
& Distills cloud intents into preference vectors to safely guide edge RL. \\

\cite{sun2026large} 
& 2026 
& Intent-Driven 
& Dual-Expert Agent 
& Energy-efficient sub-band allocation 
& Vert. 
& LLM agent 
& Decouples instruction parsing and policy generation using dual expert models. \\

\bottomrule
\end{tabular}

\begin{tablenotes}
    \footnotesize
    \item[] \textbf{Arch.} (Architecture): \textbf{Cen.} (Centralized), \textbf{Vert.} (Vertical/Hierarchical), \textbf{Hor.} (Horizontal/Distributed), \textbf{Distill.} (Distillation), \textbf{HRL} (Hierarchical Reinforcement Learning), \textbf{MA} (Multi-Agent).
\end{tablenotes}
\end{threeparttable}

\end{table*}

\subsection{Wireless-Native Foundation Models for Resource Allocation}\label{subsec:Wirless-FM4RA}

While cross-domain adaptation is effective, the fundamental modality gap between discrete linguistic spaces and continuous wireless signals often limits representational capacity. 
Inspired by the scaling laws and pre-training paradigms prevalent in LLM development, a more principled approach is to design and train wireless-native foundation models. By explicitly pre-training on massive domain-specific datasets, such as channel states and interference graphs, these compact and specialized models can bypass modality mismatches and provide a general-purpose engine for diverse resource allocation tasks. Formally, the wireless interference network can be modeled as a fully connected directed graph $\mathcal{G} = (\mathcal{V}, \mathcal{E})$, where each node $v_k \in \mathcal{V}$ represents a direct communication link, and each directed edge $e_{kj} \in \mathcal{E}$ characterizes the interference link.

Existing graph pre-training strategies generally follow generative or contrastive paradigms. Generative paradigms, such as GraphMAE \cite{hou2022graphmae}, typically focus on reconstructing node features or predicting discrete link existence. These objectives, however, are misaligned with wireless problems, where the most informative relational structure is carried by continuous edge features representing interference strength rather than simple binary connectivity. Conversely, contrastive methods like GraphCL \cite{you2020graph} rely on negative sampling, which is computationally prohibitive and conceptually difficult in the fully connected interference graphs of ultra-dense networks.

To overcome these architectural and pre-training limitations, the graph foundation model for resource allocation (GFM-RA) \cite{sheng2026gfmra} has been proposed as a pioneering solution. Unlike conventional task-specific solvers, GFM-RA builds foundational capabilities through self-supervised pre-training across a heterogeneous continuum of network densities and interference topologies, extracting a universal and robust representation space of the wireless environment. GFM-RA employs an interference-aware Transformer backbone that explicitly injects continuous physical edge features into attention scores via a bias projector, ensuring that the model's global reasoning is strictly grounded in the physics of wireless interference. This architecture is trained via a hybrid strategy, synergizing masked edge prediction to capture local link correlations with a negative-free Teacher-Student contrastive objective to guarantee global representation consistency against topological perturbations. Consequently, GFM-RA enables rapid, sample-efficient adaptation to diverse downstream utilities, ranging from sum-rate maximization to strict proportional fairness, even in out-of-distribution scenarios.

\subsection{Agentic Foundation Models for Resource Allocation}\label{subsec:Agentic-FM4RA}


Real-world resource orchestration is highly dynamic and complex, and effective resource management demands moving beyond static, isolated solvers to autonomous decision-making mechanisms capable of interpreting contexts, adapting to evolving states, and coordinating across multiple timescales.
Agentic LLMs have emerged as a promising new paradigm to meet these requirements. By combining contextual reasoning, external tool use, closed-loop feedback, and multi-agent interaction, they can support the generation and refinement of resource allocation strategies under complex system constraints~\cite{noh2025adaptive,Tong2025wirelessagent}. In addition, agentic LLMs are increasingly being integrated with reinforcement learning (RL), where they can complement conventional learning pipelines through feature abstraction, reward design, and high-level policy interpretation for high-dimensional and non-stationary wireless environments~\cite{zheng2026rl_opt}. Existing studies in this direction can be broadly organized into two main categories based on their primary optimization objectives: metric-driven resource optimization and intent-driven resource optimization. Table \ref{tab:agentic_fm_resource_allocation} provides a comprehensive summary of these representative works, highlighting their specific objectives, methodologies, architectural designs, and key contributions.

To pursue metric-driven resource optimization, a major research direction explores the use of agentic LLMs as powerful enhancement engines for conventional centralized and hierarchical control frameworks. In this paradigm, LLMs do not alter fundamental communication objectives, such as throughput maximization and latency minimization, but rather improve state representation, feature abstraction, and cross-timescale coordination. A representative example is the hierarchical RAN intelligent control (RIC) framework, in which a resource-intensive LLM operates in the non-real-time (Non-RT) RIC to analyze global telemetry and generate strategic guidance, while a near-real-time (Near-RT) RIC relies on lightweight RL modules to execute fast resource scheduling tasks, including transmit power allocation~\cite{bao2026llm}. Similarly, agentic LLMs can utilize RAG to improve state representations for RL-based resource control in highly dynamic open radio access network (O-RAN) environments. In this setting, domain-specific language models generate structured prompts capturing localized network states, guiding RL agents toward more precise QoS-aware scheduling decisions~\cite{lotfi2025orangouide}.

\begin{figure}[t]
\centerline{\includegraphics[width=0.92\linewidth]{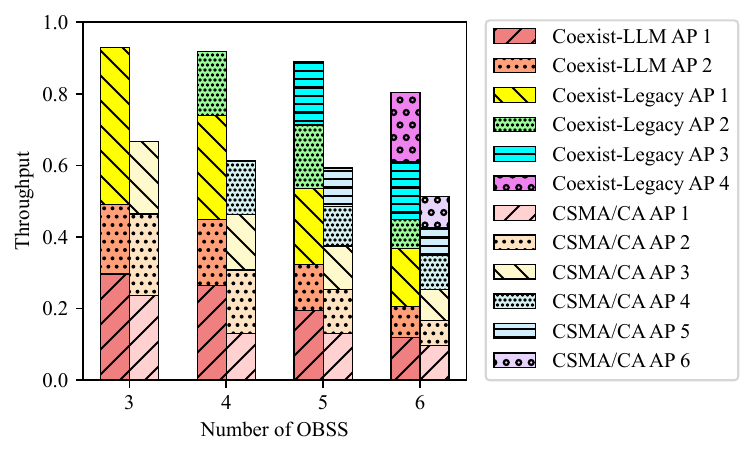}}
\caption{Coexistence between LLM-driven APs and legacy CSMA/CA APs in dense Wi-Fi networks.}
\label{legacy}
\end{figure}

Within the same objective of metric-driven resource optimization, agentic LLMs are also being explored for distributed control through multi-agent collaboration. This direction is particularly important in dense wireless deployments, where fully centralized orchestration may incur excessive signaling overhead, added latency, and limited scalability. One representative form of such collaboration is horizontal coordination among distributed wireless nodes. For example, multi-agent LLM optimization frameworks deploy separate LLM agents at different transmitters, allowing them to coordinate through contextual interaction. By sharing historical action-reward information in-context, distributed agents can gradually converge toward effective system-level power control strategies without requiring explicit mathematical models or global CSI~\cite{lee2025aidriven}. A similar horizontal collaboration paradigm has also been explored for multi-access point (AP) coordination in dense Wi-Fi networks, where each AP is modeled as an autonomous LLM agent that negotiates transmission schedules through natural-language dialogue, memory, reflection, and tool use, thereby enabling adaptive coordination under dynamic interference conditions~\cite{Fan2025mapc}. As detailed in Table~\ref{wifi6compare}, this agent-driven protocol consistently outperforms the conventional Wi-Fi 6 baseline across various network topologies. The performance improvement is mainly attributed to the proposed LLM-driven slot-level coordination mechanism, which dynamically adjusts concurrent transmission and transmission avoidance behaviors instead of relying on fixed spatial reuse parameters. Additionally, Fig.~\ref{legacy} illustrates the strong backward compatibility of this framework in practical mixed-network scenarios. Localized coordination among a few agentic APs can seamlessly coexist with legacy APs following carrier sensing multiple access with collision avoidance (CSMA/CA). By circumventing traditional back-off delays, the agent-driven devices secure system-level throughput gains while strictly preserving the normal transmissions of surrounding legacy networks.

\begin{table}[t]
\centering
\caption{Comparison of normalized throughput between Wi-Fi 6 spatial reuse and agentic protocol across various topologies}
\label{wifi6compare}
\begin{tabular}{l c c}
\toprule
\textbf{Topology} & \textbf{Wi-Fi 6} & \textbf{Agentic Protocol} \\
\midrule
\textbf{2-AP High Interference Scenario} & 0.84 & 1.00 \\
\textbf{2-AP Low Interference Scenario}   & 1.05 & 1.85 \\
\textbf{3-AP High Interference Scenario} & 0.86 & 0.91 \\
\textbf{3-AP Low Interference Scenario}   & 1.35 & 2.06 \\
\bottomrule
\end{tabular}
\end{table}

A second major category shifts the focus from predefined mathematical objectives to intent-driven resource optimization, where high-level user or operator intents are translated into executable scheduling and allocation logic. In this paradigm, the role of agentic LLMs lies not only in solving numerical optimization problems, but also in connecting semantic intent with low-level resource control. For instance, intent translation can be achieved through role-specialized multi-agent architectures, such as systems comprising separate intent and action LLM modules for interactive bandwidth allocation across multiple base stations (BSs)~\cite{han2026multillm}.

To execute these high-level intents reliably, recent works have proposed comprehensive intent-aware workflows and safeguarding mechanisms. A representative example is ComAgent, a multi-LLM framework in which specialized agents collaboratively transform natural-language requests into a solvable and verifiable optimization workflow by parsing user intents, formulating the corresponding mathematical problem, generating solver code, and validating the feasibility of the beamforming strategy before deployment~\cite{li2026comagent}. To further safeguard against unreliable actions, intent awareness can also be incorporated earlier in the optimization loop. In LAMeTA, intent-oriented knowledge distillation transfers intent-understanding capability from large cloud-based models to lightweight edge models, generating preference vectors to steer downstream RL safely~\cite{liu2026lameta}. Related approaches further include dual-expert architectures with separate instruction and policy models~\cite{sun2026large}.

\section{Emerging Technologies}
\label{sec:Emerging_Technologies}

These applications are selected because they represent four complementary dimensions of next-generation wireless intelligence: perception of the environment, reconfiguration of the electromagnetic interface, meaning-level transmission, and autonomous network control.

\subsection{Integrated Sensing and Communications}

ISAC is expected to transform future wireless networks from pure data pipes into distributed sensors of the physical
world. In an ISAC system, the same spectrum, waveform, antenna array, and network infrastructure can be reused for both information delivery and environmental perception \cite{zhang2025intelligent}. Traditional ISAC methodologies, such as localization \cite{pan2026AIdriven,xu2026Enhanced} and environment reconstruction \cite{zhang2026ER}, typically rely on simplified geometric models or task-specific NNs. However, these classical approaches often struggle to learn the underlying electromagnetic characteristics, which are governed by intrinsic wave properties and multifaceted environmental interactions, including spatial geometry and physical scattering effects.
Conventionally, the wireless channel can be mathematically characterized as the superposition of multiple propagation paths, represented as
\begin{equation}
\mathbf{H} = \sum_{l=1}^{L} \beta_l e^{-j 2 \pi f \tau_l} \mathbf{a}_{\mathrm{r}}(\theta_l)\mathbf{a}_{\mathrm{t}}^H(\phi_l),
\end{equation}
where $f$ is the carrier frequency, $L$ denotes the number of propagation paths, 
$\beta_l$ represents the complex path gain influenced by physical scattering and reflection, 
and $\tau_l$ is the propagation delay of the $l$-th path. 
In addition, $\theta_l$ and $\phi_l$ denote the AoA and AoD, respectively, while 
$\mathbf{a}_{\mathrm{r}}(\theta_l)$ and $\mathbf{a}_{\mathrm{t}}(\phi_l)$ represent the corresponding receive and transmit steering vectors. 
Traditional ISAC methodologies often struggle to accurately decompose these highly entangled EM parameters, especially in complex non-line-of-sight (NLOS) environments.

The introduction of foundation models offers a transformative paradigm shift, leveraging robust representation learning to map noisy RF data into structured latent spaces. By implicitly mastering these intricate electromagnetic interactions, foundation models facilitate a transition from basic signal sensing to profound environmental semantic understanding, providing a robust framework for the high-fidelity inference of both the geometric architecture and dynamic characteristics of the physical world. This representation-oriented view provides a natural bridge from foundation models to ISAC applications. During pretraining, the model learns how location, motion, object presence, environmental structure, and propagation conditions jointly affect received signals across time, frequency, and antenna dimensions. Once such a general representation is obtained, downstream ISAC tasks can be formulated as different readouts of the same physical-world representation.

As summarized in Table~\ref{tab:isac}, the current landscape of foundation models for ISAC can be categorized into two primary evolutionary paths: pre-trained foundation models and wireless-native foundation models. Pre-trained foundation models for ISAC treat wireless signals as a specialized language (e.g., using models such as GPT-2, LLaMA, or vision-based architectures), enabling specific ISAC tasks through fine-tuning or multi-modal alignment. For instance, MMSense in \cite{li2025MMSense} adapts the feature extraction capabilities of LVMs to interpret RF signals, bridging the gap between visual semantics and RF signal patterns. Furthermore, the high-level reasoning and generative capabilities of large models are harnessed in \cite{peng2025Large}, where a LLaMA3-driven framework orchestrates distributed integrated multi-modal sensing by fusing RF data with camera feeds to enhance semantic awareness. The primary advantage of pre-trained foundation models lies in their ability to inherit sophisticated reasoning and cross-modal association capabilities from massive-scale textual or visual pre-training. However, since their foundational knowledge is derived from non-electromagnetic modalities, they often struggle to achieve a deep, physics-aware understanding of intrinsic electromagnetic characteristics.

\begin{table*}[t]
\centering
\caption{Summary of Existing Foundation Models for ISAC}
\label{tab:isac}
\renewcommand{\arraystretch}{1.2}
\setlength{\tabcolsep}{3pt}

\begin{tabular}{ccc >{\centering\arraybackslash}p{2cm} c p{2.8cm} p{5.2cm}}
\toprule
Reference & Year & Model & Type & Multi-modal  & Modalities & Key Contribution \\
\midrule

\cite{li2025MMSense} 
& 2025 
& MMSense 
& Pre-trained foundation model
& \ding{51} 
& Image, RF, LiDAR, Text 
& Unified multi-modal framework for joint channel, human, and environment sensing \\

\cite{peng2025Large} 
& 2025 
& LLM-DiSAC 
& Pre-trained foundation model
& \ding{51} 
& RF, Image 
& LLaMA3-driven distributed framework that combines RF-vision fusion \\

\cite{yazdnian2026mmISAC} 
& 2026 
& MM-ISAC 
& Wireless-native foundation model
& \ding{51}
& RF, 3D Occupancy Grid
& Physics-guided foundation model for cross-modal EM propagation learning \\

\cite{Henk2025LWLM} 
& 2025 
& LWLM
& Wireless-native foundation model
& \ding{55} 
& RF
&  Hybrid pre-training tasks combining masked modeling, domain invariance, and contrastive learning for wireless localization  \\

\cite{zhu2026amfm} 
& 2026 
& AM-FM
& Wireless-native foundation model
& \ding{51} 
& RF
& Physics-informed foundation model overcoming device heterogeneity for diverse WiFi sensing tasks \\

\cite{cheng2026wifom2}
& 2026 
& WiFo-M$^2$
& Wireless-native foundation model
& \ding{51} 
& RF, LiDAR, Image
& Contrastive foundation model aligning multi-modal with CSI for plug-and-play physical action enhancement across diverse tasks \\

\bottomrule
\end{tabular}
\end{table*}

To overcome these domain-specific limitations, wireless-native foundation models for ISAC operate directly on wireless observations, such as CSI, I/Q samples, and CIRs, rather than relying on semantic representations transferred from other modalities. These observations inherently reflect the interactions between wireless propagation and the physical environment, carrying rich information about multipath propagation, Doppler effects, scattering, object locations, and motion. Through self-supervised pre-training on massive and heterogeneous wireless datasets, wireless-native foundation models can learn universal representations that capture both channel characteristics and environment-dependent propagation patterns. Such representations can then be adapted to a wide range of ISAC downstream tasks, including communication tasks such as channel prediction and beam management, as well as sensing tasks such as localization, target detection and tracking, activity recognition, and environment reconstruction.

For example, the foundation model developed in \cite{yazdnian2026mmISAC} adopts a physics-guided self-supervised architecture that processes raw CSI using 3D environmental occupancy grids and user locations as auxiliary structural constraints. Alongside masked reconstruction, it introduces a global aggregator token to reconstruct the spatial spectrum. This pre-training enables the model to identify underlying electromagnetic propagation phenomena within the CSI. By explicitly anchoring the latent space to wave propagation physics, the model extracts highly transferable, environment-agnostic representations, allowing the backbone to be seamlessly fine-tuned for diverse ISAC downstream tasks with strong zero-shot and few-shot generalization in unseen physical environments. The large wireless localization model (LWLM) \cite{Henk2025LWLM} directly extracts localization-relevant semantics from raw CSI. By employing a hybrid self-supervised learning framework that integrates spatial-frequency masked channel modeling, domain-transformation invariance, and contrastive learning, LWLM enables various ISAC downstream inferences, such as time-of-arrival (ToA), AoA, single and multi-BS positioning.
Extending this paradigm to ubiquitous WiFi sensing, AM-FM \cite{zhu2026amfm} pre-trains on massive in-the-wild CSI data using contrastive learning, masked reconstruction, and a physics-informed autocorrelation objectives to capture the dynamics. By addressing hardware and environmental heterogeneity, this unified backbone enables highly efficient few-shot transfer across diverse ambient intelligence applications, ranging from physiological monitoring to spatial tracking.

Building upon both types of foundation models, the frontier of ISAC is rapidly evolving toward multi-modal foundation models. While wireless signals offer unique advantages like all-weather operability and NLOS sensing, they inherently suffer from lower spatial resolution compared to optical sensors. Acting as universal feature extractors with exceptional representation and reasoning capabilities, foundation models naturally support the integration of heterogeneous multi-modal data. For instance, the framework in \cite{cheng2026wifom2} fuses RF signals, images, and LiDAR point clouds within a shared latent space, effectively mitigating the inherent perceptual blind spots of unimodal sensing. Concurrently, this multi-modal fusion elevates ISAC beyond mere geometric perception to a comprehensive semantic intelligence, thereby providing a powerful paradigm for advanced 6G applications.

\subsection{Fluid Antenna Systems and New MIMO Systems}

Fluid antenna systems (FASs) and other emerging MIMO architectures, such as near-field MIMO, extremely large-scale MIMO (XL-MIMO), movable antennas, holographic surfaces, and tri-hybrid MIMO, are reshaping the physical-layer design space of 6G networks~\cite{heath2025tri2,heath2025tri,New2025FASTutorial}. Unlike conventional fixed-array MIMO, these systems introduce reconfigurable spatial degrees of freedom by allowing antenna positions, radiation patterns, visibility regions, and electromagnetic apertures to be dynamically adapted to the propagation environment. This flexibility, however, also creates a fundamentally more complex design problem: the channel is no longer determined only by the environment and transceiver locations, but also by the configuration of the antenna aperture itself. Consequently, key operations such as port selection, beamforming, channel prediction, antenna allocation, and user scheduling become high-dimensional, nonlinear, and often combinatorial. Foundation models provide a promising methodology for addressing this complexity by learning reusable representations of channel evolution, spatial correlation, and configuration-dependent electromagnetic responses across diverse deployment scenarios.

A first research direction adapts pre-trained LLMs to FAS channel and port prediction~\cite{fas_llm_empowered_Wnag}. In FAS-enabled mobile links, the optimal port may change rapidly due to user mobility, Doppler effects, and strong spatial correlation among neighboring ports. The Port-LLM framework in~\cite{fas_llm_empowered_Wnag} addresses this issue by converting historical channel tables across movable ports into sequential inputs and then using a LoRA-adapted pre-trained LLM to predict future channel tables and corresponding moving ports. This two-step formulation, channel-table forecasting followed by port selection, allows the FAS to proactively relocate the active port so that the effective channel remains approximately stable over time. The same work further introduces Prompt-Port-LLM, a prompt-enhanced variant that incorporates dynamic prompts and a trainable prompt encoder to inject real-time statistical information, such as the mean, variance, and extrema of the input channel data, into the LLM adaptation process. These designs illustrate how an LLM originally trained for text can be repurposed as a sequence predictor for wireless dynamics, provided that specialized embedding and projection modules are introduced to bridge the modality gap between continuous complex-valued CSI and discrete language tokens.

Foundation models are also being explored for high-mobility and non-terrestrial FAS links. In OTFS-enabled satellite-FAS systems, the channel evolves jointly over the spatial-port and delay-Doppler domains, making direct prediction prohibitively high-dimensional. FAS-LLM in~\cite{fas_llm_Yang} tackles this challenge through a two-stage compression strategy that first selects representative reference-port information and then applies separable principal component analysis (PCA) across spatial and delay-Doppler dimensions. The resulting compact representations are embedded into a LoRA-adapted LLM for multi-step channel forecasting. This approach demonstrates an important design principle for wireless foundation models: rather than feeding raw high-dimensional channel tensors directly into an LLM, physics-aware compression can expose the dominant channel structure while reducing the input dimension to a scale compatible with foundation-model inference. Such delay-Doppler-aware forecasting is particularly relevant for FAS-assisted satellite IoT and other fast-varying links, where proactive port reconfiguration and link adaptation must be completed before the predicted channel becomes outdated.

Beyond prediction, LLMs can serve as general-purpose optimizers for joint FAS configuration and transceiver design. In multi-user MISO-FAS systems, port selection and beamforming are strongly coupled: a port configuration that appears favorable under one beamformer may become suboptimal once power allocation and multi-user interference are jointly considered. The LLM-based framework in \cite{fas_port_llm_Guo} departs from conventional sequential pipelines that first select ports and then optimize beamforming. Instead, it uses a pre-trained LLM backbone, LoRA fine-tuning, and task-specific pre- and post-processing modules to output port indices and beamforming-related variables in parallel. Differentiable relaxations, such as the Gumbel-Sinkhorn mechanism, further allow discrete port-selection decisions to be incorporated into end-to-end training. This line of work suggests that the multi-task representation capacity of foundation models can be exploited not merely for prediction, but also for directly mapping CSI to coupled physical-layer control decisions.

Another emerging direction is the use of LLMs as hyper-heuristic designers for FAS optimization. Since joint port selection and precoder design typically leads to mixed-integer nonconvex optimization, classical algorithms often rely on manually crafted heuristics, alternating optimization, or repeated convex relaxations. The LLM-driven design framework in \cite{fas_port_llm_Wang} introduces a hyper-heuristic perspective, where LLMs are used to generate, critique, and refine heuristic operators for combinatorial search. For example, ReEvo-style frameworks employ one LLM to generate genetic-algorithm crossover operators and another LLM to reflect on their performance, producing “verbal gradients” that guide subsequent heuristic evolution. This paradigm is particularly attractive for FAS because the optimal search strategy may depend on the number of ports, aperture size, user density, signal-to-interference-plus-noise ratio (SINR) constraints, and channel correlation structure. By automating the design of heuristics, LLMs can reduce dependence on human-crafted rules and enable adaptive optimization across heterogeneous FAS deployments.

The role of foundation models naturally extends from FAS to broader next-generation MIMO systems. In near-field and XL-MIMO systems, the channel exhibits spherical wavefronts, spatial non-stationarity, user-specific visibility regions, and distance-angle coupling, which challenge far-field beamforming and conventional CSI compression. Wireless-native foundation models trained directly on large-scale RF datasets could learn universal propagation representations that capture these near-field structures across carrier frequencies, array geometries, and environments. Meanwhile, LVMs and multi-modal foundation models can fuse camera, LiDAR, radar, map, and RF measurements to infer blockage, user location, scatterer geometry, and visibility regions, thereby supporting proactive beam focusing and antenna-subarray activation. In this sense, FAS, near-field MIMO, XL-MIMO, and tri-hybrid MIMO share a common foundation-model opportunity: the physical aperture becomes an adaptive interface whose configuration can be predicted, optimized, and reasoned about using learned representations of the surrounding radio environment.

Despite this promise, several challenges remain before foundation models can be deployed in practical FAS and new MIMO systems. Table~\ref{tab:open_challenges} summarizes representative challenges that must be addressed to advance wireless foundation models from signal-level representation learning toward reliable network-level autonomy. First, the modality gap between complex-valued wireless signals and language or vision tokens requires principled tokenization, embedding, and normalization strategies. Second, real-time port switching and beam adaptation impose stringent latency and energy constraints, motivating lightweight wireless-native models, small language models, and edge-deployable adapters rather than monolithic cloud-scale LLMs. Third, robustness and physical consistency must be guaranteed under distribution shifts caused by mobility, hardware impairments, mutual coupling, calibration errors, and unseen propagation environments. Finally, future foundation-model-based FAS design should move from task-specific demonstrations—such as port prediction, channel forecasting, joint port selection and beamforming, and LLM-generated heuristics—toward unified models that jointly support channel prediction, port selection, beamforming, sensing, and cross-layer control. Such models could ultimately transform reconfigurable MIMO architectures from static hardware extensions into intelligent, self-optimizing electromagnetic systems.

\subsection{Semantic Communication}
To accommodate the explosive data demands of bandwidth-intensive, delay-sensitive applications such as holographic telepresence and autonomous driving, semantic communication has emerged as a powerful complement to traditional bit-exact transmission~\cite{BeyondBits}. Semantic communication fundamentally shifts the design focus from the isolated, symbol-by-symbol replication of data to the end-to-end delivery of underlying meaning and intent. Pre-trained foundation models, including LLMs/LVMs, act as effective catalysts for this shift. By serving as high-quality semantic knowledge bases (SKBs) and integrated semantic codecs, foundation models enable a holistic, joint-design philosophy that leverages vast pre-trained world knowledge to significantly enhance system-wide transmission efficiency.

Moving beyond independent source and channel coding, foundation models are redefining semantic communication through a unified semantic representation pipeline. Traditional deep joint source-channel coding (JSCC) circumvents the digital cliff effect by mapping source data directly into continuous-valued symbols. Foundation models elevate this approach by aligning NLP pipelines with physical-layer objectives. For example, the LLM-SC framework utilizes an LLM's native tokenizer as a semantic encoder, enabling robust text transmission at low SNRs without explicit channel coding \cite{LLMSC}. For dense visual data, architectures such as LaMoSC facilitate multi-modal fusion via cross-attention between visual streams and textual prompts \cite{LaMoSC}. This allows receivers to logically infer and reconstruct visual geometries, even those that are severely degraded by channel fading, using robust semantic priors.
Along this line, prompt-based transmission achieves ultra-low-rate communication by substituting raw data with concise textual prompts. In this architecture, VLMs compress a complex visual signal into a text prompt, which subsequently drives a generative decoding process to reconstruct the original content. To ensure precise visual control alongside this text-driven generation, hybrid approaches have been proposed that preserve fidelity by segmenting images based on their semantic importance \cite{Prompt2025GLOBECOM}. Critical areas undergo precise, image-oriented feature extraction, while non-critical backgrounds are abstracted into structured text descriptors. The receiver then seamlessly merges these transmitted visual features with text-anchored synthesis, bypassing classical rate-distortion limits through VLM-driven contextual generation. Furthermore, this hybrid extraction strategy is highly resilient in severe propagation environments. Building on this concept, foundation model-empowered satellite networks leverage dynamic semantic segmentation to overcome physical challenges such as high mobility, interference, and long delays. For instance, the FMSAT framework in \cite{Jiang2025SemanticSatellite} selectively protects critical visual features and leverages cross-image correlation for adaptive, diffusion-based reconstruction, avoiding costly delayed retransmissions and guaranteeing robust semantic delivery.

Building upon these generative reconstruction strategies, semantic communication systems are leveraging integrated AI-generated content engines that treat the noisy communication channel as an active part of the reconstruction process. Advanced diffusion-driven semantic communication frameworks exploit the wireless environment by directly mapping stochastic channel noise into the forward diffusion process \cite{Diff2025SemCom}. Guided by extracted semantic priors, the reverse denoising phase systematically removes channel-induced interference, synthesizing photorealistic content from ultra-low-rate anchors. 
However, such generative reconstruction may still suffer from semantic inconsistency or hallucinated details when the received information is highly incomplete. To address this issue, \cite{tang2025retrieval} incorporates RAG into GenAI-enabled semantic communication, allowing the receiver to retrieve relevant knowledge from historical transmissions or external semantic repositories as additional grounding information. By combining retrieval with diffusion-based semantic image reconstruction, the framework improves semantic consistency, enhances adaptability to diverse tasks and dynamic environments, and reduces the risk of generating content that deviates from the intended semantics.

\begin{table*}[t]
\centering
\caption{Open Challenges and Possible Directions for Wireless Foundation Models}
\label{tab:open_challenges}
\renewcommand{\arraystretch}{1.18}
\begin{tabular}{p{0.20\textwidth} p{0.36\textwidth} p{0.36\textwidth}}
\toprule
\textbf{Challenge} & \textbf{Why It Matters} & \textbf{Possible Directions} \\
\midrule
EM-domain modality gap 
& Wireless signals are continuous, complex-valued, and non-stationary. 
& Physics-informed tokenization and complex-valued embeddings. \\

Real-time deployment 
& PHY/RAN timescales impose strict latency and energy constraints. 
& Lightweight foundation models, adapters, and edge inference. \\

Data scarcity and benchmarking 
& Real-world wireless data are difficult to collect, synchronize, and share. 
& Digital twins, federated datasets, and standardized benchmarks. \\

Trustworthy agentic AI 
& LLM agents may hallucinate or violate communication constraints. 
& Formal verification, constrained reasoning, and tool-grounded control. \\

Cross-layer integration 
& PHY, resource-management, and network-control tasks are tightly coupled. 
& Unified wireless-native models and multi-agent orchestration. \\
\bottomrule
\end{tabular}
\end{table*}
\vspace{-3mm}
\subsection{Agentic LLMs for Autonomous Network Management}

The evolution toward next-generation networks necessitates a transition from traditional manual network configurations to fully autonomous, closed-loop network management. In this context, the integration of agentic LLMs has emerged as a key enabler, serving as intelligent assistants throughout the network lifecycle. Agentic LLMs extend network intelligence from interpreting complex technical standards and operational knowledge, to supporting network control and optimization under existing system rules, and ultimately to assisting in the design of novel communication protocols~\cite{lira2025large}. At the foundational level, agentic LLMs can help interpret complex technical standards and operational knowledge, thereby providing the semantic grounding needed for subsequent decision-making. Building on this understanding, agentic LLMs can further support network control and optimization by translating high-level natural-language intents into actionable configuration and orchestration decisions. At a more advanced level, their capabilities may further extend beyond operation under existing rules toward supporting the design of new communication protocols.

The hierarchical complexity and multi-modal nature of standards documents, such as those released by the 3rd Generation Partnership Project (3GPP), pose significant challenges for LLM-based automated processing. To address this issue, recent research has introduced domain-specific resources, such as the TSpec-LLM dataset~\cite{nikbakht2024tspec}, which encompasses 3GPP releases 8 through 19 and provides a foundational corpus for pre-training and fine-tuning network-centric models. Building upon such foundations, advanced agentic RAG frameworks, such as TelcoAI~\cite{ghosh2025telcoai}, have been proposed to seamlessly fuse textual and visual information. This multi-modal integration significantly enhances semantic retrieval and contextual understanding for network operation tasks.

Beyond standards interpretation and knowledge retrieval, agentic LLMs are increasingly deployed for zero-touch configuration and hyperparameter optimization. In highly dynamic environments, including O-RAN architectures, the tuning of Layer-1 to Layer-3 parameters requires rapid adaptation to real-time network telemetry. Foundational models such as ORANSight-2.0 have been specifically fine-tuned using parameter-efficient methods tailored for O-RAN architectures, thereby enabling automated parameter adjustments that substantially outperform general-purpose commercial models~\cite{gajjar2025oransight}. In addition to parameter tuning, agentic LLMs have also been extended to broader operational functions, including network security and analytics. For example, multi-agent frameworks that integrate LLMs directly into these functions have been developed~\cite{liu2025toward}. These frameworks allow agents to dynamically translate high-level operator intents into secure, verifiable infrastructure configurations while systematically mitigating potential vulnerabilities. Related research in general multi-agent systems, such as SentinelAgent~\cite{he2025sentinelagent}, investigates runtime oversight through dynamic agent--tool interaction graphs, node-, edge-, and path-level anomaly detection, and policy-based intervention, offering a perspective potentially relevant to agent-assisted network management.

At the protocol-design level, agentic LLMs are also being explored for communication protocol design, particularly in emerging semantic communication frameworks. In such settings, structured inter-agent interactions enable more selective and demand-aware information exchange, thereby reducing redundant transmissions and improving bandwidth efficiency. For instance, frameworks employing retrieval-augmented multi-modal semantic perception allow autonomous agents, such as connected vehicles, to initially exchange only task-critical semantics and retrieve high-resolution data patches only when needed. This semantics-aware communication protocol reduces redundant transmissions while helping satisfy stringent latency constraints, and it highlights the potential of agentic LLMs to enable new protocol primitives for semantics-aware, on-demand wireless communication~\cite{liu2026wireless}. 

Despite rapid progress, many existing studies remain proof-of-concept demonstrations. Several works adapt general-purpose LLMs to wireless tasks through tokenization or prompting, but do not yet demonstrate large-scale pre-training on diverse wireless data. Others show transfer across limited scenarios but lack systematic evaluation under hardware impairments, mobility, topology changes, or real-time constraints. Therefore, a key open question is how to define and evaluate foundation-model behavior in wireless systems beyond model size or architectural similarity to LLMs.

\section{Challenges and Opportunities}
\label{sec:challenges}

This section outlines key challenges and promising research directions required to fully unlease the potential of foundation models in designing next-generation wireless communication systems.

\subsection{Physics-Informed Representation Learning for Wireless Signals}
Raw wireless signals are highly unstructured, noisy, and strictly governed by complex electromagnetic wave propagation. The application of foundation models to the physical layer requires extracting universal, physics-aware representations from these signals while adhering to strict microsecond-level latency constraints. Without explicitly anchoring the model’s latent space to intrinsic physical propagation rules, foundation models tend to overfit domain-specific statistical artifacts. Meanwhile, to meet latency requirements, conventional compression techniques (e.g., quantization, pruning, and distillation) are often applied, but they inevitably degrade the model's performance. Consequently, over-compressed or physics-agnostic models collapse into narrow, scenario-specific networks, losing the essential zero-shot capabilities. Future research shall develop lightweight, edge-friendly architectures capable of bridging this gap. Achieving this near-optimal trade-off requires exploring self-supervised physics-informed representation learning, dynamic early-exit inference, and hybrid designs that integrate signal processing algorithms with lightweight neural reasoning.

\vspace{-1mm}
\subsection{Expanding the Frontier of Resource Decisions}
While the feasibility of wireless-native foundation models is established, foundation models for wireless resource decision remains in their infancy, primarily addressing simplified problems like decentralized power control and user scheduling. Moving forward, these models must naturally expand to handle more complex resource dimensions, including dynamic spectrum access, multi-user scheduling, and MIMO beamforming. Tackling these high-dimensional, often NP-hard problems requires advancing architectural designs, by developing novel pre-training strategies capable of capturing global network states and non-stationary temporal dynamics over large-scale interference topologies.
Beyond offline pre-training, another important direction is to establish a closed-loop learning paradigm in which operational wireless data are continuously collected during network deployment, used to update and adapt the foundation model online, and subsequently translated into improved resource decisions and network performance. Such an online cycle of data collection, learning, and network improvement will be essential for enabling foundation models to continuously adapt to evolving wireless environments and resource demands.

\subsection{Unified Datasets and Standardized Benchmarks}
Wireless foundation models require large-scale, high-quality datasets that capture the distinctive characteristics of wireless systems. For physical-layer models, training data should contain high-fidelity wireless observations collected across diverse propagation environments, frequency bands, antenna configurations, mobility conditions, and hardware settings. For resource allocation, datasets should further include interference topologies, traffic patterns, channel dynamics, and service requirements. However, existing datasets are often proprietary or fragmented, making standardized and openly accessible wireless datasets essential for large-scale pre-training and model generalization.

Equally important is the establishment of standardized benchmarks for evaluating and comparing different models. They should provide consistent data splits, tasks, metrics, and test-isolation protocols, while assessing generalization, latency, robustness, and adaptation under unseen conditions. Recently, CFM-Bench \cite{gao2026cfm} represents an initial effort in this direction. Beyond task-specific performance, future benchmarks should also evaluate whether learned representations can transfer across PHY, RAN, and ISAC, supporting tasks ranging from channel inference and RAN optimization to localization, tracking, and environment reconstruction.

\subsection{Safety, Verification, and Latency of Agentic Foundation Models}
While agentic foundation models enable advanced automated network management, issues like hallucinations, catastrophic forgetting, lack of domain knowledge may result in network configurations that seem correct but are actually flawed or inefficient. This creates severe reliability risks in practice. Current safety methods, such as basic rule checks, can block simple errors but cannot ensure the correctness of complex, multi-step plans when hallucinations occur. Therefore, it is important to investigate sophisticated safety verification mechanisms to ensure that the generated plan meets strict wireless rules before execution to guarantee safe model behavior. Furthermore, relying on lengthy natural language reasoning causes long processing delays. Developing agentic models that maintain high-level planning skills but output decisions using compact symbols or vector embeddings represents another direction for meeting strict real-time requirements.

\section{Conclusion}
In this paper, we have provided a comprehensive overview of the emerging landscape of foundation models for wireless communications, tracing their evolution from signal-level representation learning to reasoning-driven network autonomy. First, off-the-shelf pre-trained foundation models have been adapted to wireless tasks, enabling data-efficient transfer across channel estimation, prediction, detection, beam management, and resource allocation. Second, wireless-native foundation models have emerged to bridge the modality gap between discrete semantic tokens and continuous wireless signals by pre-training directly on wireless data and utilizing compact domain knowledge. Third, agentic foundation models have extended static representation learning toward autonomous, reasoning-driven network orchestration. Beyond these core operations, we also discussed how foundation models can support emerging 6G frontiers, including ISAC, fluid antenna systems and new MIMO architectures, semantic communications, and system-level network autonomy. Despite this promise, realizing the full potential of foundation models in wireless networks requires overcoming several critical bottlenecks. Addressing these challenges will pave the way toward fully intelligent, adaptive, and autonomous 6G networks.

\small
\bibliographystyle{IEEEtran}
\bibliography{ref.bib}

\end{document}